\documentclass[a4paper,hyper,cits]{PoS-vaso}
\usepackage{url,units,xspace}
\usepackage{paralist}

\usepackage{amsmath,amsfonts,amssymb,amsthm}
\allowdisplaybreaks
\usepackage{slashed,extarrows}
\usepackage{commath}
\usepackage{tikz-feynman, contour}

\usepackage{mathrsfs}
\usepackage{graphicx,subcaption} 
\usepackage{enumitem}
\usepackage{float,ragged2e,multirow}

\DeclareGraphicsExtensions{.pdf,.png,.jpg}
\graphicspath{{./}{./figs/}}

\newcommand{\MAD}{\mbox{\textsc{MadGraph}}\xspace}
\newcommand{\Feyn}{\textsc{FeynRules}\xspace}
\newcommand{\Math}{\textsc{Mathematica}\xspace}
\newcommand{\pyt}{\textsc{Python}\xspace}
\newcommand{\fort}{\textsc{Fortran}\xspace}

\newcommand{\gd}{\ensuremath{g_{\textrm D}}\xspace}

\newcommand{\sgg}{s_{\gamma\gamma}\xspace}
\newcommand{\half}{\nicefrac{1}{2}\xspace}
\newcommand{\dd}{\ensuremath{\mathrm d}}
\newcommand{\pt}{\ensuremath{p_{\mathrm T}}\xspace}

\title{Monopole production via photon fusion at the LHC }

\ShortTitle{Monopole production via photon fusion at the LHC }

\author{\speaker{Vasiliki A.\ Mitsou} \\ 
        Instituto de F\'isica Corpuscular (IFIC), CSIC -- Universitat de Val\`encia, \\ 
C/ Catedr\'atico Jos\'e Beltr\'an 2, E-46980 Paterna (Valencia), Spain\\
        E-mail: \email{vasiliki.mitsou@ific.uv.es}}

\abstract{The existence of magnetic monopoles, also predicted in some GUT theories, would symmetrise Maxwell equations and explain the charge quantisation. Searches for them are being performed in cosmic telescopes as well as in collider experiments, such as MoEDAL and ATLAS. In this report, we focus on the, least explored, photon-fusion mechanism, yet Drell--Yan results will be presented, too. Cross sections for monopoles of spin~0,~\half and~1 for an effective monopole-velocity-dependent magnetic charge are presented. For spin-\half and spin-1 monopoles, a magnetic-moment term is included, which is treated as a new phenomenological parameter and, together with the velocity-dependent coupling, allows for a perturbative treatment of the cross-section calculation. We present an appropriate implementation of photon-fusion and Drell--Yan processes into \MAD UFO models, aimed to serve as a useful tool in monopole searches at LHC, especially for photon fusion, given that it has not been considered by experimental collaborations recently. Moreover, the experimental implications of such perturbatively reliable monopole searches are discussed.  }

\FullConference{Corfu Summer Institute 2018 "School and Workshops on Elementary Particle Physics and Gravity"\\
		(CORFU2018)\\
		31 August - 28 September, 2018\\
		Corfu, Greece}

\begin{document}

\section{Introduction}\label{intro}

The magnetic monopole remains a hypothetical particle, years since its concrete formulation by Dirac~\cite{dirac} as a quantum mechanical source of magnetic poles. Although there are concrete field-theoretical models beyond the Standard Model (SM) of particle physics which contain concrete monopole solutions~\cite{hpmono,cho,you,aruna,sarben,shafi}, these are extended objects with substructure, and their production at colliders is either impossible, as they are ultra-heavy~\cite{hpmono,shafi}, or extremely suppressed, due to their composite nature~\cite{drukier}. On the other hand, Dirac point-like monopoles are sources of singular magnetic fields, the underlying quantum field theory of which, if any, is completely unknown. 

In ref.~\cite{Baines:2018ltl}, the results of which are highlighted here, effective field theoretical models to obtain production mechanisms were considered based on \emph{electric-magnetic duality}, i.e.\ on the replacement of the electric charge $q_e$ by the magnetic charge $g$, the latter obeying Dirac's quantisation rule:
\begin{equation}\label{diracrule}
g q_e = 2\pi n, \quad n \in {\mathbb Z}.
\end{equation}
The fine-structure constant is given by $\alpha =  \frac{e^2}{4\pi} = \frac{1}{137}$ with $e$ the positron charge, hence~\eqref{diracrule} yields
\begin{equation}\label{diracrule2}
g = \frac{1}{2\alpha} n  \, \Big(\frac{e}{q_e} \Big)\, e  = 68.5e \, \Big(\frac{e}{q_e}\Big) \, n \equiv n \,  \frac{e}{q_e}\, \gd, \quad n \in {\mathbb Z},
\end{equation}
with $\gd = 68.5e$ the fundamental Dirac charge. This electric-magnetic duality replacement may be used as a basis for the evaluation of monopole-production cross sections from collisions of SM particles. Unfortunately, due to the large value of the magnetic charge~\eqref{diracrule2}, such a replacement renders the corresponding production process non-perturbative. Nevertheless, one may attempt to set benchmark scenarios for extracting mass limits by using Feynman-like graphs. This is standard practice in all monopole searches at colliders so far~\cite{mm}. 

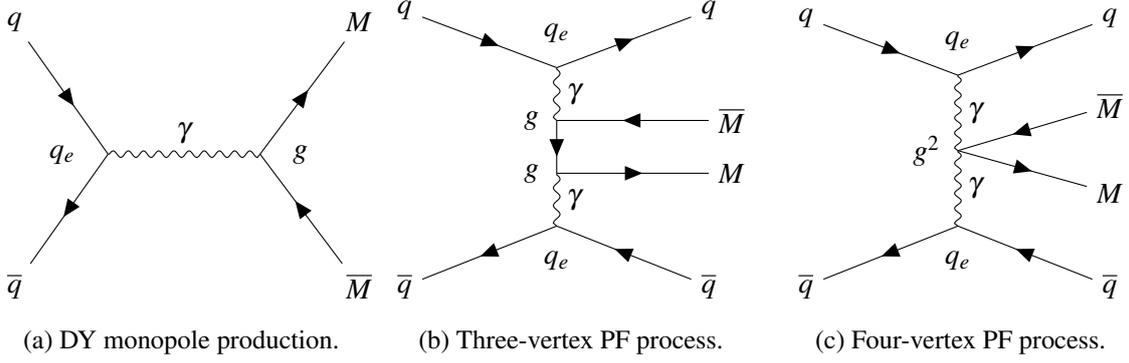
\begin{figure}[ht!]
\begin{subfigure}[b]{0.32\textwidth}
\centering
\begin{tikzpicture}
  \begin{feynman}
  	\vertex (a1);
	\vertex[left=0.3cm of a1] (a11)  {\(q_e\)};
	\vertex[right=2cm of a1] (a2) ;
	\vertex[right=0.3cm of a2] (a22)  {\(g\)};
	\vertex[above=1.75cm of a2] (a3);
	\vertex[right=1cm of a3] (a4) {\(M\)};
	\vertex[below=1.75cm of a2] (a5);
	\vertex[right=1cm of a5] (a6) {\(\overline{M}\)};
	
	\vertex[above=1.75cm of a1] (a7);
	\vertex[left=1cm of a7] (a8) {\(q\)};
	\vertex[below=1.75cm of a1] (a9);
	\vertex[left=1cm of a9] (a10) {\(\overline{q}\)};
	
	\diagram* {
	       	(a8) -- [fermion] (a1) -- [fermion] (a10),
		(a1) -- [boson, edge label=\(\gamma\)] (a2),
       		(a6) -- [fermion] (a2) -- [fermion] (a4),
      };
\end{feynman}
\end{tikzpicture}
\caption{DY monopole production.\label{fig:dy}}
\end{subfigure}
~
\begin{subfigure}[b]{0.32\textwidth}
\centering
\begin{tikzpicture}
  \begin{feynman}
  	\vertex (a1);
	\vertex[above=0.2cm of a1] (a11)  {\(q_e\)};
	\vertex[right=2cm of a1] (a2);
	\vertex[above=0.5cm of a2] (a3)  {\(q\)};
	\vertex[left=2cm of a1] (a4);
	\vertex[above=0.5cm of a4] (a5)  {\(q\)};

	\vertex[below=2.1cm of a1] (a7);	
	\vertex[below=0.2cm of a7] (a22)  {\(q_e\)};
	\vertex[right=2cm of a7] (a8);
	\vertex[below=0.5cm of a8] (a9) {\(\overline{q}\)};
	\vertex[left=2cm of a7] (a81);
	\vertex[below=0.5cm of a81] (a91) {\(\overline{q}\)};
	
	\vertex[below=0.7cm of a1] (a71);	
	\vertex[left=0.1cm of a71] (a711) {\(g\)};	
	\vertex[below=1.4cm of a1] (a72);	
	\vertex[left=0.1cm of a72] (a722) {\(g\)};	
	
	\vertex[right=2cm of a71] (a82) {\(\overline{M}\)};
	\vertex[right=2cm of a72] (a83) {\(M\)};		
	
	\diagram* {
	       	(a5) -- [fermion] (a1) -- [fermion] (a3),
		(a9) -- [fermion] (a7) -- [fermion] (a91),
		(a1) -- [boson, edge label=\(\gamma\)] (a71),
		(a72) -- [boson, edge label=\(\gamma\)] (a7),
		(a82) -- [fermion] (a71) -- [fermion] (a72) -- [fermion] (a83),
      };
\end{feynman}
\end{tikzpicture}
\caption{Three-vertex PF process.\label{fig:pf3}}
\end{subfigure}
~
\begin{subfigure}[b]{0.32\textwidth}
\centering
\begin{tikzpicture}
  \begin{feynman}
  	\vertex (a1);
	\vertex[above=0.2cm of a1] (a11)  {\(q_e\)};
	\vertex[right=2cm of a1] (a2);
	\vertex[above=0.5cm of a2] (a3)  {\(q\)};
	\vertex[left=2cm of a1] (a4);
	\vertex[above=0.5cm of a4] (a5)  {\(q\)};

	\vertex[below=2cm of a1] (a7);	
	\vertex[below=0.2cm of a7] (a22)  {\(q_e\)};
	\vertex[right=2cm of a7] (a8);
	\vertex[below=0.5cm of a8] (a9) {\(\overline{q}\)};
	\vertex[left=2cm of a7] (a81);
	\vertex[below=0.5cm of a81] (a91) {\(\overline{q}\)};
	
	\vertex[below=1cm of a1] (a17);
	\vertex[left=0.1cm of a17] (a171) {\(g^{2}\)};
	
	\vertex[right=2cm of a17] (a82);
	\vertex[above=0.3cm of a82] (a821) {\(\overline{M}\)};
	\vertex[below=0.3cm of a82] (a822) {\(M\)};		
	
	\diagram* {
	       	(a5) -- [fermion] (a1) -- [fermion] (a3),
		(a9) -- [fermion] (a7) -- [fermion] (a91),
		(a1) -- [boson, edge label=\(\gamma\)] (a17) -- [boson, edge label=\(\gamma\)] (a7),
		(a821) -- [fermion] (a17) -- [fermion] (a822),
      };
\end{feynman}
\end{tikzpicture}
\caption{Four-vertex PF process.\label{fig:pf4}}
\end{subfigure}
\caption{Feynman-like tree-level graphs for monopole--antimonopole pair production processes. \subref{fig:dy} DY monopole production via $q\bar{q}$ annihilation; \subref{fig:pf3} monopole production via photon fusion (spins 0, \half and 1); \subref{fig:pf4} contact diagram for monopole production via photon-fusion (spins~0 and~1). }
\label{fig:diagrams}
\end{figure}

Depending on the spin of the monopole field, typical graphs of monopole--antimonopole pair production at LHC from proton--proton ($pp$) collisions are given in fig.~\ref{fig:diagrams}. There are two kinds of such processes: the Drell--Yan (DY) (see fig.~\ref{fig:dy}) and the photon-fusion (PF) 
 production (see figs.~\ref{fig:pf3} and~\ref{fig:pf4}).  In ref.~\cite{Baines:2018ltl}, previous DY-plus-PF studies~\cite{original,dw} carried out for spin-\half monopoles were extended to also consider monopoles of spin~$0$ and~1. The most recent MoEDAL analysis~\cite{Acharya:2019vtb} makes use of  the results on photon-fusion production of ref.~\cite{Baines:2018ltl} presented here.

An arbitrary value for the magnetic dipole moment for monopoles with spin is discussed, treating $\kappa$ as a \emph{new phenomenological parameter}. A  large-$\kappa$ limit may allow, despite the large magnetic coupling, the monopole-pair production cross sections to be finite, thus making sense of the Feynman-like graphs of the effective theory. These perturbativity conditions~\cite{Baines:2018ltl}, which pertain to slowly moving monopoles, are of relevance to MoEDAL searches~\cite{moedal-review}.

This report is structured as follows. In section~\ref{sec:spin}, the cross sections for monopole--antimonopole pair production via PF are derived for various monopole spins. The pertinent Feynman rules are implemented in a dedicated \MAD model, described in section~\ref{sec:mad}, together with the monopole phenomenology at the LHC. Conclusions and outlook are given in section~\ref{sec:concl}.  

\section{Cross sections for spin-$S$ monopole production via photon--photon scattering \label{sec:spin}}

We discuss the electromagnetic interaction of a monopole of spin $S=0,~\half,~1$ with a photon. The corresponding theory is an effective 
$U(1)$ gauge theory obtained after appropriate \emph{dualisation} of the pertinent field theories describing the interactions of charged spin-$S$ fields with photons. 
Milton, Schwinger and collaborators conjectured~\cite{milton,sch}, upon invoking electric--magnetic duality, that when discussing the interaction of monopole with matter (electrons or quarks), e.g.\ the propagation of monopoles in matter media used for detection and capture of monopoles, or  considering monopole--antimonopole pair production through DY or PF processes, a \emph{monopole-velocity dependent magnetic charge} has to be considered in the corresponding cross section formulae:
\begin{equation}\label{gbeta}
g\,  \rightarrow \,  g \, \frac{v}{c} \equiv g\, \beta .
\end{equation}

The replacement \eqref{gbeta} was then used to interpret the  experimental data in collider searches for magnetic monopoles~\cite{mm,original,dw,milton,kalb,vento,atlasmono1,moedalplb}. Due to the lack of a concrete theory for magnetic sources, the results of the pertinent experimental searches can be interpreted in terms of {\it both} a $\beta$-independent and a $\beta$-dependent magnetic charges, and then one may compare the corresponding bounds, as in the recent searches by the MoEDAL Collaboration~\cite{Acharya:2019vtb,moedalplb}. The monopole velocity $\beta$ is given by the \emph{Lorentz invariant expression} in terms of the monopole mass $M$ and the Mandelstam variable $s$, related to the centre-of-mass energy of the fusing incoming particles, i.e.\ photons or (anti)quarks ($\sqrt{s}=2E_{\gamma/q}$):
\begin{equation}\label{btos}
 \beta = \sqrt{1-\frac{4M^2}{s}}~.
 \end{equation}
As a result of~\eqref{gbeta}, the coupling of the monopole to the photon, $g(\beta)$, is linearly dependent on the particle boost, $\beta= |\vec{p}| / E_p$, where $(E_p,\vec{p})$ is the monopole four-momentum. 

In this section we outline the pertinent Feynman rules and then proceed to give  expressions for the associated differential and total production cross sections
for the photon-fusion production for monopole fields of various spins. A detailed discussion on the Drell--Yan process can be found in ref.~\cite{Baines:2018ltl}. We consider both $\beta$-dependent and $\beta$-independent magnetic couplings. 

\subsection{Scalar monopole}

A massive spin-0 monopole interacting with a massless $U(1)$ gauge field representing the photon is discussed here. Searches at the LHC~\cite{atlasmono1,moedalplb,atlasmono2,moedal} and other colliders~\cite{mm} have set upper cross-section limits is this scenario assuming DY production. The Lagrangian describing the electromagnetic interactions of the monopole is given simply by a dualisation of the SQED Lagrangian
\begin{equation}\label{spinzero}
\mathcal{L}=-\frac{1}{4}F^{\mu\nu}F_{\mu\nu}+(D^{\mu}\phi)^{\dagger}(D_{\mu}\phi)-M^2\phi^{\dagger}\phi ,
\end{equation}
where $D_{\mu}=\partial_{\mu}-ig(\beta)\mathcal{A}_{\mu}$, $\mathcal{A}_{\mu}$ is the photon field with strength tensor $F_{\mu\nu}=\partial_{\mu}\mathcal{A}_{\nu}-\partial_{\nu}\mathcal{A}_{\mu}$ and $\phi$ is the scalar monopole field. There are two interaction vertices, that couple the spin-0 monopole with the $U(1)$ gauge field. 

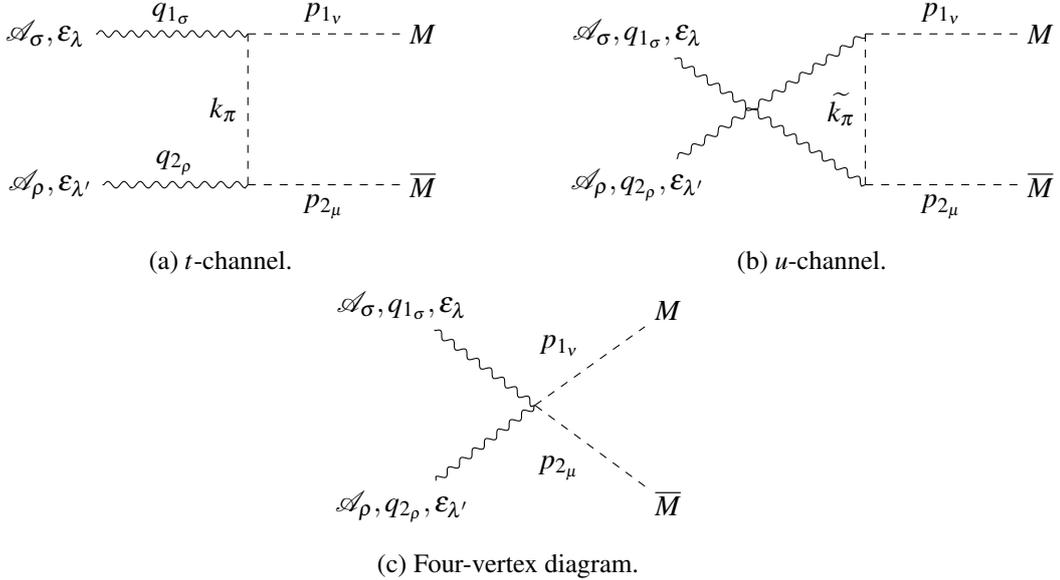
\begin{figure}[ht!]
\begin{subfigure}[b]{0.5\textwidth}
\centering
\begin{tikzpicture}
  \begin{feynman}
  	\vertex (a1) {\(\mathcal{A}_{\rho}, \epsilon_{\lambda'}\)};
	\vertex [right=2.6cm of a1] (a2);
	\vertex [right=2cm of a2] (a3){\(\overline{M}\)};
	\vertex [above=2cm of a2] (a5);
	\vertex [right=2cm of a5] (a6){\(M\)};
	\vertex [left=2cm of a5] (a4){\(\mathcal{A}_{\sigma}, \epsilon_{\lambda}\)};
	\diagram* {
	(a3) -- [scalar, edge label=\(p_{2_{\mu}}\)] (a2) -- [scalar, edge label=\(k_{\pi}\)] (a5) -- [scalar, edge label=\(p_{1_{\nu}}\)] (a6),
	(a4) -- [boson, edge label=\(q_{1_{\sigma}}\)] (a5),
	(a1) -- [boson, edge label=\(q_{2_{\rho}}\)] (a2)
         };
\end{feynman}
\end{tikzpicture}
\caption{$t$-channel.\label{fig:tchan}}
\end{subfigure}
~
\begin{subfigure}[b]{0.5\textwidth}
\centering
\begin{tikzpicture}
  \begin{feynman}
  	\vertex (a1) {\(\mathcal{A}_{\rho} ,q_{2_{\rho}},  \epsilon_{\lambda'}\)};
	\vertex [right=3cm of a1] (a2);
	\vertex [right=2cm of a2] (a3){\(\overline{M}\)};
	\vertex [above=2cm of a2] (a5);
	\vertex [right=2cm of a5] (a6){\(M\)};
	\vertex [left=2cm of a5] (a4){\(\mathcal{A}_{\sigma} ,q_{1_{\sigma}},\epsilon_{\lambda}\)};
	\diagram* {
	(a3) -- [scalar, edge label=\(p_{2_{\mu}}\)] (a2) -- [scalar, edge label=\(\widetilde{k_{\pi}}\)] (a5) -- [scalar, edge label=\(p_{1_{\nu}}\)] (a6),
	(a4) -- [boson] (a2),
	(a1) -- [boson] (a5)
      }; 
\end{feynman}
\end{tikzpicture}
\caption{$u$-channel.\label{fig:uchan}}
\end{subfigure}
~
\begin{subfigure}[b]{1\textwidth}
\centering
\begin{tikzpicture}
\begin{feynman}
  	\vertex (a2);
	\vertex[left=1.75cm of a2] (a3);
	\vertex[below=1cm of a3] (a4) {\(\mathcal{A}_{\rho} , q_{2_{\rho}}, \epsilon_{\lambda'}\)};
	\vertex[right=1.75cm of a2] (a5);
	\vertex[below=1cm of a5] (a6) {\(\overline{M}\)};
	\vertex[above=1cm of a3] (a8) {\(\mathcal{A}_{\sigma} ,q_{1_{\sigma}},  \epsilon_{\lambda}\)};
	\vertex[above=1cm of a5] (a7) {\(M\)};
	\diagram* {
       		(a6) -- [scalar, edge label=\(p_{2_{\mu}}\)] (a2) -- [scalar, edge label=\(p_{1_{\nu}}\)] (a7),
		(a8) -- [boson] (a2) -- [boson] (a4)
      };
\end{feynman}
\end{tikzpicture}
\caption{Four-vertex diagram.\label{fig:seagull}}
\end{subfigure}
\caption{Feynman-like graphs for: \subref{fig:tchan} $t$-channel;  \subref{fig:uchan} $u$-channel; and \subref{fig:seagull} seagull processes encompass all the contributions to the matrix amplitude of scalar particle production by PF.}
\label{Spinzerographs}
\end{figure}

There are three possible graphs contributing to scalar monopole production by PF: $t$-channel, $u$-channel and seagull graph shown in fig.~\ref{Spinzerographs}. The differential cross section for the spin-0 monopole--antimonopole production in terms of the pseudorapidity $\eta$ yields~\cite{Baines:2018ltl} 
\begin{align}\label{dsigmadomegazerorap}
	\frac{\dd\sigma_{\gamma\gamma\rightarrow M\overline{M}}^{S=0}}{\dd\eta}  = \frac{\pi\alpha_{g}^{2}(\beta)\beta}{\sgg \cosh^{2}\eta} \left[ 1+\left(1-\frac{2(1-\beta^{2})}{1-\beta^{2}\tanh^{2}\eta}\right)^{2} \right].
\end{align}
The pair production is mostly central as is the case for various beyond-SM scenarios. After integrating over the solid angle, the total cross section reads
\begin{equation} \label{tcrosssection}
	\sigma^{S=0}_{\gamma\gamma\rightarrow M\overline{M}} =\frac{4\pi\alpha_{g}^{2}(\beta)\beta}{\sgg}\left[2-\beta^{2}-\frac{1}{2\beta}(1-\beta^{4})\ln{\left(\frac{1+\beta}{1-\beta}\right)}\right].
\end{equation}
The cross section drops rapidly with increasing mass and disappears sharply at the kinematically forbidden limit of $M > \sqrt{\sgg}/2$.  

\subsection{Spin-\half point-like monopole with arbitrary magnetic moment term}

The phenomenology of a monopole with spin~\half is the most thoroughly studied case~\cite{original,dw,kalb,vento} and explored in collider searches~\cite{Acharya:2019vtb,atlasmono1,moedalplb,atlasmono2,moedal}. The electromagnetic interactions of the monopole with photons, are described by a $U(1)$ gauge theory for a spinor field $\psi$ representing the monopole interacting with the massless $U(1)$ gauge field $A^{\mu}$ of the photon.  

The unknown origin of the monopole magnetic moment leads us to assume that it is generated only through anomalous quantum-level spin interactions as for the QED electron. If $\kappa=0$, the Dirac Lagrangian is recovered. Thus, the effective Lagrangian for the spinor-monopole-photon interactions takes the form
\begin{equation}\label{12lag}
\mathcal{L}=-\frac{1}{4}F_{\mu\nu}F^{\mu\nu}+\overline{\psi}(i\slashed{D}-m)\psi-i\frac{1}{4}\, g(\beta)\, \kappa \, F_{\mu\nu}\overline{\psi}[\gamma^{\mu},\gamma^{\nu}]\psi,
\end{equation}
where $F_{\mu\nu}$ is the electromagnetic field strength tensor, $\slashed{D}=\gamma^{\mu}(\partial_{\mu}-ig(\beta)\mathcal{A}_{\mu})$ is the total derivative and $[\gamma^{\mu},\gamma^{\nu}]$ is a commutator of $\gamma$ matrices. 
The magnetic coupling $g(\beta)$  is at most linearly dependent on the monopole boost $\beta$. 
The effect of the magnetic-moment term is observable through its influence on the magnetic moment at tree level which is
\begin{equation}
\mu_{M}=\frac{g(\beta)}{2M}2(1+2\tilde{\kappa})\hat{S}, \quad \hat{S}=\frac{1}{2},
\end{equation}
where $\hat{S}$ is the spin expectation value and the corresponding ``gyromagnetic ratio'' $g_R = 2( 1 + 2\tilde{\kappa})$. The dimensionless constant $\tilde{\kappa}$ is defined such that
\begin{equation}\label{ktilde}
\kappa=\frac{\tilde{\kappa}}{M}. 
\end{equation}

Noticeably, the parameter $\kappa$ in \eqref{12lag}  is dimensionful having units of inverse mass, which breaks the renormalisability of the theory. This may not be a serious obstacle for considering the case $\kappa\neq0$, if the pertinent model is considered in the context of an effective field theory embedded in some, yet unknown, microscopic theory, in which the renormalisation can be recovered.

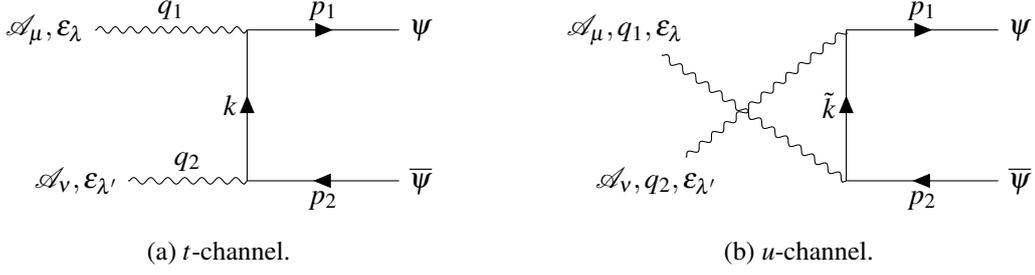
\begin{figure}[ht!]
\begin{subfigure}[b]{0.5\textwidth}
\centering
\begin{tikzpicture}
  \begin{feynman}
  	\vertex (a1) {\(\mathcal{A}_{\nu}, \epsilon_{\lambda'}\)};
	\vertex [right=2.25cm of a1] (a2);
	\vertex [right=2cm of a2] (a3){\(\overline{\psi}\)};
	\vertex [above=2cm of a2] (a5);
	\vertex [right=2cm of a5] (a6){\(\psi\)};
	\vertex [left=2cm of a5] (a4){\(\mathcal{A}_{\mu}, \epsilon_{\lambda}\)};
	\diagram* {
	(a3) -- [fermion, edge label=\(p_{2}\)] (a2) -- [fermion, edge label=\(k\)] (a5) -- [fermion, edge label=\(p_{1}\)] (a6),
	(a4) -- [boson, edge label=\(q_{1}\)] (a5),
	(a1) -- [boson, edge label=\(q_{2}\)] (a2)
         };
\end{feynman}
\end{tikzpicture}
\caption{$t$-channel.\label{fig:tchan-pf-half}}
\end{subfigure}
\begin{subfigure}[b]{0.5\textwidth}
\centering
\begin{tikzpicture}
  \begin{feynman}
  	\vertex (a1) {\(\mathcal{A}_{\nu},  q_{2},  \epsilon_{\lambda'}\)};
	\vertex [right=2.5cm of a1] (a2);
	\vertex [right=2cm of a2] (a3){\(\overline{\psi}\)};
	\vertex [above=2cm of a2] (a5);
	\vertex [right=2cm of a5] (a6){\(\psi\)};
	\vertex [left=2cm of a5] (a4){\(\mathcal{A}_{\mu} ,  q_{1},  \epsilon_{\lambda}\)};
	\diagram* {
	(a3) -- [fermion, edge label=\(p_{2}\)] (a2) -- [fermion, edge label=\(\tilde{k}\)] (a5) -- [fermion, edge label=\(p_{1}\)] (a6),
	(a4) -- [boson] (a2),
	(a1) -- [boson] (a5)
      }; 
\end{feynman}
\end{tikzpicture}
\caption{$u$-channel.\label{fig:uchan-pf-half}}
\end{subfigure}
\caption{Feynman-like graphs for the $t$-channel~\subref{fig:tchan-pf-half} and $u$-channel~\subref{fig:uchan-pf-half} show the contributions to the matrix amplitude for pair production of spin-\half monopoles by PF. }
\label{spinhalfgraphs}
\end{figure}

\begin{figure}[ht!]\centering
\includegraphics[width=\textwidth]{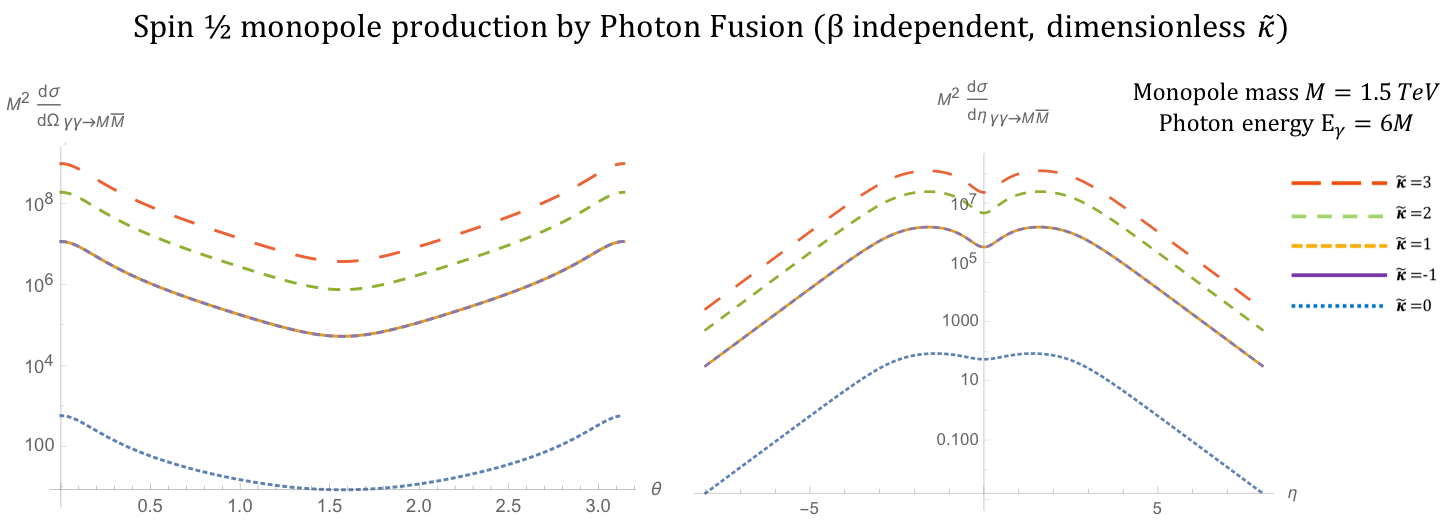}
\caption{Angular distributions for monopole--antimonopole pair production via PF for a monopole with spin~\half and mass $M=1.5$~TeV as a function of the scattering angle $\theta$ (left) and the pseudorapidity $\eta$ (right) at $\sqrt{\sgg}=2E_{\gamma}$, where $E_{\gamma}=6M$, and for various values of the parameter $\tilde{\kappa}$~\cite{Baines:2018ltl}.  }\label{LeeYangDsigDstuff12label}
\end{figure}

Photon fusion occurs through $t$-channel and $u$-channel processes as shown in fig.~\ref{spinhalfgraphs}. The angular distributions are shown in fig.~\ref{LeeYangDsigDstuff12label} for spin~\half and for various values of the parameter $\tilde{\kappa}$.   The case $\tilde{\kappa}=0$ represents the SM expectation for electron-positron pair production if the coupling is substituted by the electric charge $e$, i.e.\ restoring the Lagrangian to simple Dirac QED. This case is clearly distinctive as the only unitary and renormalisable case~\cite{Baines:2018ltl}. 

\begin{figure}[ht]
\includegraphics[width=0.55\linewidth]{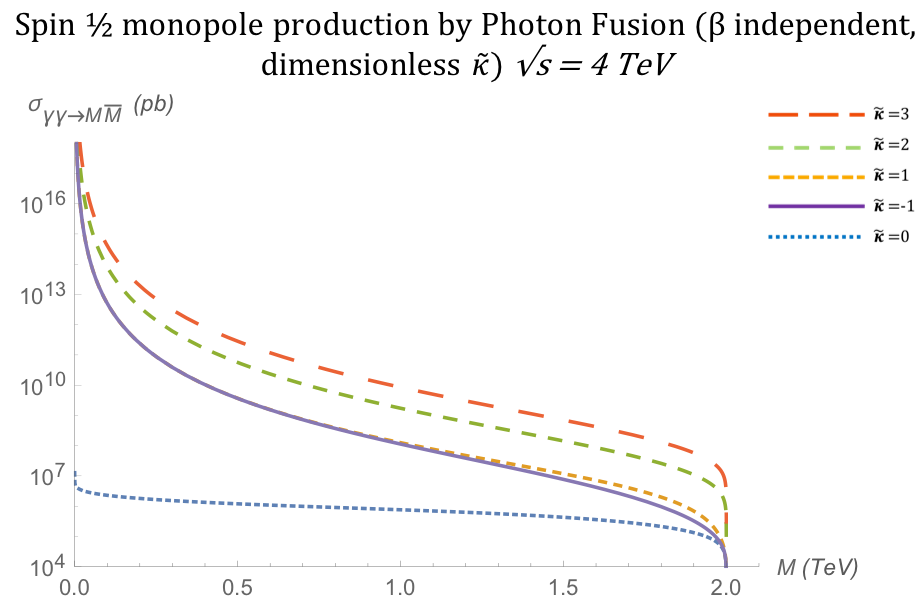}\hspace{0.05\linewidth}%
\begin{minipage}[b]{0.4\linewidth}\caption{\label{LeeYangsig12label}   Total cross section for the pair production of spin-\half monopoles via the PF process, as a function of the monopole mass $M$ for different values of $\tilde{\kappa}$ at $\sqrt{\sgg}=4$~TeV~\cite{Baines:2018ltl}.  }
\end{minipage}
\end{figure}

The total cross section for arbitrary $\kappa$ is shown graphically in fig.~\ref{LeeYangsig12label}.
The high-energy limit $\sgg\rightarrow \infty$ makes the differential cross section divergent, except, unsurprisingly, when recovering the Dirac model with $\kappa=0$~\cite{Baines:2018ltl}: 
\begin{align}
 	\frac{\dd\sigma^{S=\frac{1}{2}}_{\gamma\gamma\rightarrow M\overline{M}}}{\dd\Omega} \quad \xrightarrow{\sgg\rightarrow \infty} &\quad  \frac{\alpha_g^2(\beta)}{\sgg} \frac{1 + \cos ^2 \theta}{1-\cos^2 \theta}, \quad\quad \kappa=0.
 \end{align}
 The total cross section carries the same high energy behaviour; it is finite for $\kappa=0$, and diverges for all other values of $\kappa$:
\begin{equation}\label{txs}
	\sigma^{S=\frac{1}{2}}_{\gamma\gamma\rightarrow M\overline{M}} \quad \xrightarrow{\sgg\rightarrow \infty} \quad  
	\begin{cases}
		\sgg &, \quad\quad \kappa\neq 0, \nonumber \\
		\dfrac{4 \pi  \alpha_g^2(\beta) }{\sgg} \Big[ \ln \Big(\dfrac{\sgg}{M^2}\Big) -1 \Big] &, \quad\quad \kappa=0.  
	\end{cases}
\end{equation}
For a constant value of $\sgg$, as assumed in fig.~\ref{LeeYangsig12label}, the high-energy limit may be approximated by $M \rightarrow 0$, where the cross section becomes finite only for the value $\kappa=0$ and diverges in other $\kappa$ values, as expected from \eqref{txs}. 

Hence, a unitarity requirement may isolate the $\kappa=0$ model from the others as the only viable theory for the spin-\half monopole, unless the model is viewed as an effective field theory. In that case, the value of $\kappa$ can be used as a window to extrapolate some characteristics of the extended model in which unitarity is restored. In addition, $\kappa$ is dimensionful, hence a non-zero $\kappa$ clearly makes this (effective) theory non-renormalisable.

\subsection{Vector monopole with arbitrary magnetic moment term \label{sec:spin1}}

Monopoles of spin-1 have been addressed for the first time in colliders recently by the MoEDAL experiment for the Drell--Yan production~\cite{moedalplb}. A monopole with a spin $S=1$ is postulated as a \emph{massive} vector meson $W_{\mu}$ interacting only with a massless gauge field $\mathcal{A}_{\mu}$ in the context of a gauge invariant Proca field theory. Lacking a fundamental theory for point-like magnetic poles,  a magnetic moment term is included in the effective Lagrangian, proportional to $\kappa$, as a free phenomenological parameter~\cite{Baines:2018ltl}. Unlike the spin-\half monopole case, for the vector monopole the magnetic moment parameter $\kappa$ is dimensionless. The case $\kappa=0$ corresponds to a pure Proca Lagrangian, and $\kappa=1$ to that of the SM $W_{\mu}$ boson in a Yang--Mills theory with spontaneous symmetry breaking. As we shall see, for certain formal limits of large $\kappa$ and slowly moving monopoles, one may also attempt to make sense of the perturbative DY or PF processes of monopole--antimonopole pair production, when velocity-dependent magnetic charges are employed. 

 The pertinent effective Lagrangian, obtained by imposing electric--magnetic duality on the respective Lagrangian terms for the interaction of $W^{\pm}$ gauge bosons with photons in the generalised SM framework~\cite{Tupper} takes the form:
	\begin{align}\label{GaugeInvLag}
	& \mathcal{L}=-\xi(\partial_{\mu}W^{\dagger\mu})(\partial_{\nu}W^{\nu})-\frac{1}{2}(\partial_{\mu}\mathcal{A}_{\nu})(\partial^{\nu}\mathcal{A}_{\mu})-\frac{1}{2}G_{\mu\nu}^{\dagger}G^{\mu\nu}-M^{2}W_{\mu}^{\dagger}W^{\mu}-ig(\beta)\kappa F^{\mu\nu}W^{\dagger}_{\mu}W_{\nu}~,
	\end{align}
where $\dagger$ denotes the hermitian conjugate and $G^{\mu\nu}=(D^{\mu}W^{\nu}-D^{\nu}W^{\mu})$, with $D_{\mu}=\partial_{\mu}-ig(\beta)\, \mathcal{A}_{\mu}$ the $U(1)$ covariant derivative, which provides 
the coupling of the (magnetically charged) vector field $W_{\mu}$ to the gauge field $\mathcal{A}_{\mu}$, playing the role of the ordinary photon.  
The parameter $\xi$ is a gauge-fixing parameter. The magnetic and quadrupole moments are given respectively by
\begin{subequations}
\begin{align}
\mu_M=\frac{g({\beta})}{2M}&(1+\kappa)\hat{S}, \quad \hat{S}=1, \label{magmom}\\
Q_E&=-\frac{g(\beta)\kappa}{M^2}\label{quadmom}~,
\end{align}
\end{subequations}
where $\hat{S}$ is the monopole spin expectation value and the corresponding ``gyromagnetic ratio'' $g_R = 1 + \kappa$. Restricting our attention to tree level, \eqref{GaugeInvLag} is used to evaluate the $\kappa$-dependent PF Born amplitudes $\mathcal{A}_{\mu}\mathcal{A}_{\nu}\rightarrow W_{\mu}W^{\dagger}_{\nu}$ and DY $\psi\overline{\psi}\rightarrow W_{\mu}W^{\dagger}_{\nu}$. 

\begin{figure}[ht!]\centering
\includegraphics[width=\textwidth]{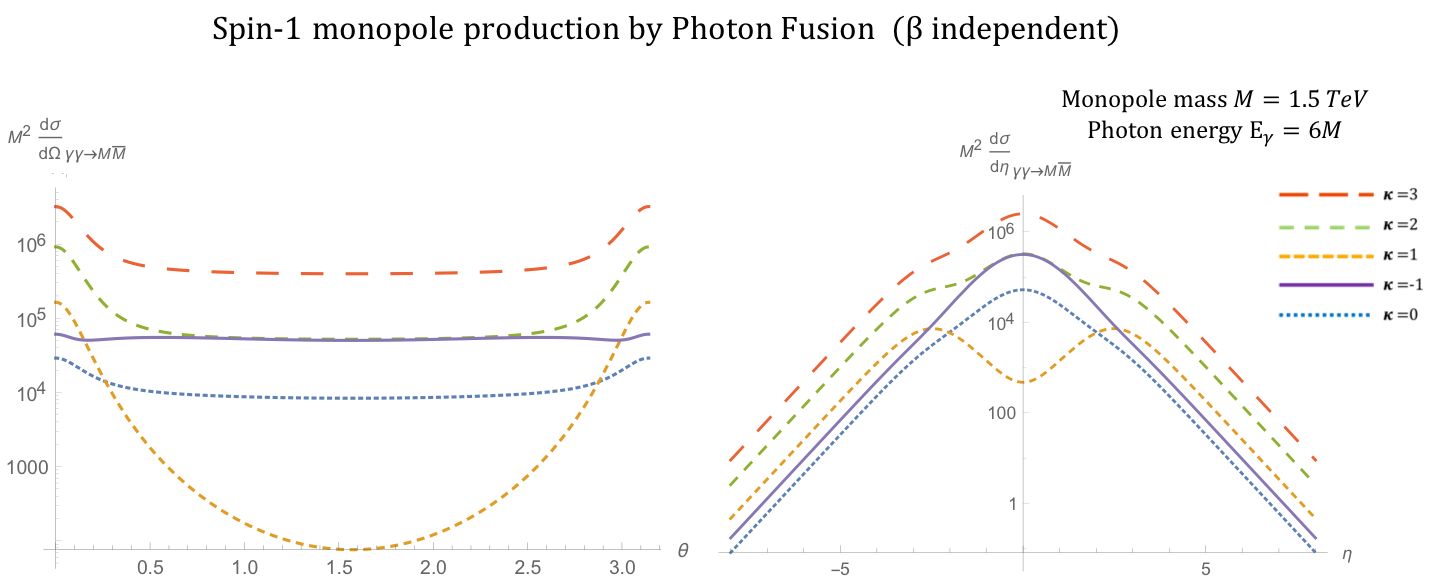}
\caption{\label{LeeYangDsigDomega1pflabel} Angular distributions for monopole--antimonopole pair production via PF for a monopole with spin~1 and mass $M=1.5$~TeV as a function of the scattering angle $\theta$ (left) and the pseudorapidity $\eta$ (right) at $\sqrt{\sgg}=2E_{\gamma}$, where $E_{\gamma}=6M$, and for various values of the parameter $\kappa$~\cite{Baines:2018ltl}. }
\end{figure}

Monopole--antimonopole pairs with spin~1 are generated at tree level by photons fusing in the $t$-channel, $u$-channel and at a 4-point vertex in  Feynman-like graphs similar to those given in fig.~\ref{Spinzerographs} for scalar monopoles~\cite{Baines:2018ltl}. These kinematic distributions are plotted as functions of the kinematic variables $\theta$ and $\eta$ in fig.~\ref{LeeYangDsigDomega1pflabel} for various values of the phenomenological parameter $\kappa$. 

The distinct behaviour of the kinematic distribution in the unitarity-respecting case $\kappa=1$, that shows a depression around $\eta=0$, as compared to the peaks in the cases where $\kappa \ne 1$ is shown in fig.~\ref{LeeYangDsigDomega1pflabel}.
This is in agreement with the situation characterising $W^+W^-$ production in the SM case~\cite{Tupper}.
The parameter $\kappa$ influences the unitarity and renormalisability of the effective theory; the $\kappa=1$ case  is the \emph{only finite} solution in the ultraviolet limit $\sgg \to \infty$~\cite{Baines:2018ltl}: 
\begin{align}
\frac{\dd\sigma^{S=1}_{\gamma\gamma\rightarrow M\overline{M}}}{\dd\Omega} \quad \xrightarrow{\sgg\rightarrow \infty} \quad  \frac{\alpha_g^2(\beta)}{2\sgg( 1-\cos^2 \theta )^2}\left(3\cos^4 \theta +10\cos^2 \theta +19\right), \quad\ \kappa=1.
\end{align}
For all other $\kappa$ values one obtains a differential cross section, which diverges linearly with $\sgg \to \infty$:
\begin{align}
\frac{\dd\sigma^{S=1}_{\gamma\gamma\rightarrow M\overline{M}}}{\dd\Omega} \quad  \xrightarrow{\sgg\rightarrow \infty} \quad  \sgg, \quad\quad \kappa\neq1.
\end{align}

\begin{figure}[ht]
\includegraphics[width=0.55\linewidth]{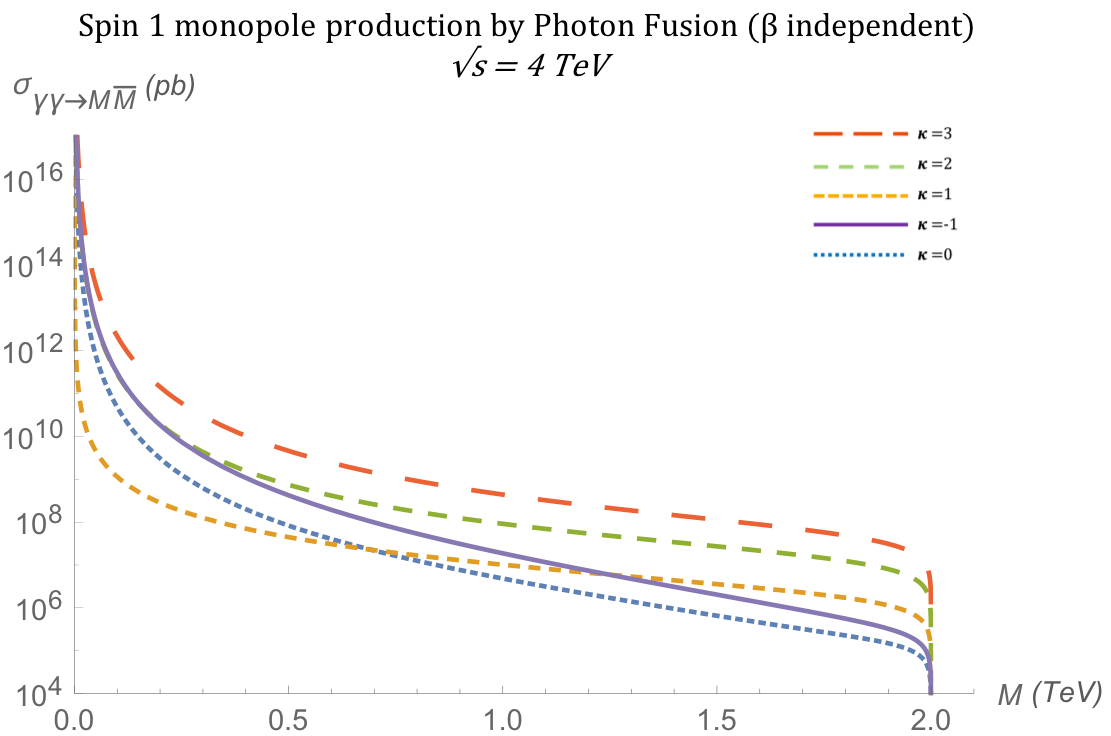}\hspace{0.05\linewidth}%
\begin{minipage}[b]{0.4\linewidth}\caption{\label{LeeYangsig1pflabel} Total cross section for the pair production of spin-1 monopoles via the PF process, as a function of the monopole mass $M$ for different values of $\kappa$ at $\sqrt{\sgg}=4$~TeV~\cite{Baines:2018ltl}.  }
\end{minipage}
\end{figure}

The total cross section for the $\kappa=1$ case is given by
\begin{equation}\label{reproRusakovich1}
\begin{split}
\sigma_{\gamma\gamma\rightarrow M\overline{M}}^{S=1}=&\frac{\pi \alpha_g^2(\beta)\beta}{\sgg}\left(2\frac{3\beta^4-9\beta^2+22}{1-\beta^2}-\frac{3(1-\beta^4)}{\beta}\ln\left(\frac{1+\beta}{1-\beta}\right)\right),
\end{split}
\end{equation}
and  is shown graphically in fig.~\ref{LeeYangsig1pflabel} for various $\kappa$ values. In the high energy limit $\sgg\rightarrow \infty$,  \emph{only} the total cross section for $\kappa=1$ is \emph{finite}, in similar spirit to the differential cross section behaviour:
\begin{equation}
\sigma^{S=1}_{\gamma\gamma\rightarrow M\overline{M}} \quad \xrightarrow{\sgg\rightarrow \infty} \quad  
\begin{cases}
\dfrac{8\pi\alpha_g^2(\beta)}{M^2} &, \quad \kappa=1, \nonumber \\ 
 \sgg &, \quad  \kappa \neq 1.
\end{cases}
\end{equation}

\subsection{Comparison of various spin models and production processes}\label{sec:comparison} 

\begin{figure}[ht!]
\includegraphics[width=\textwidth]{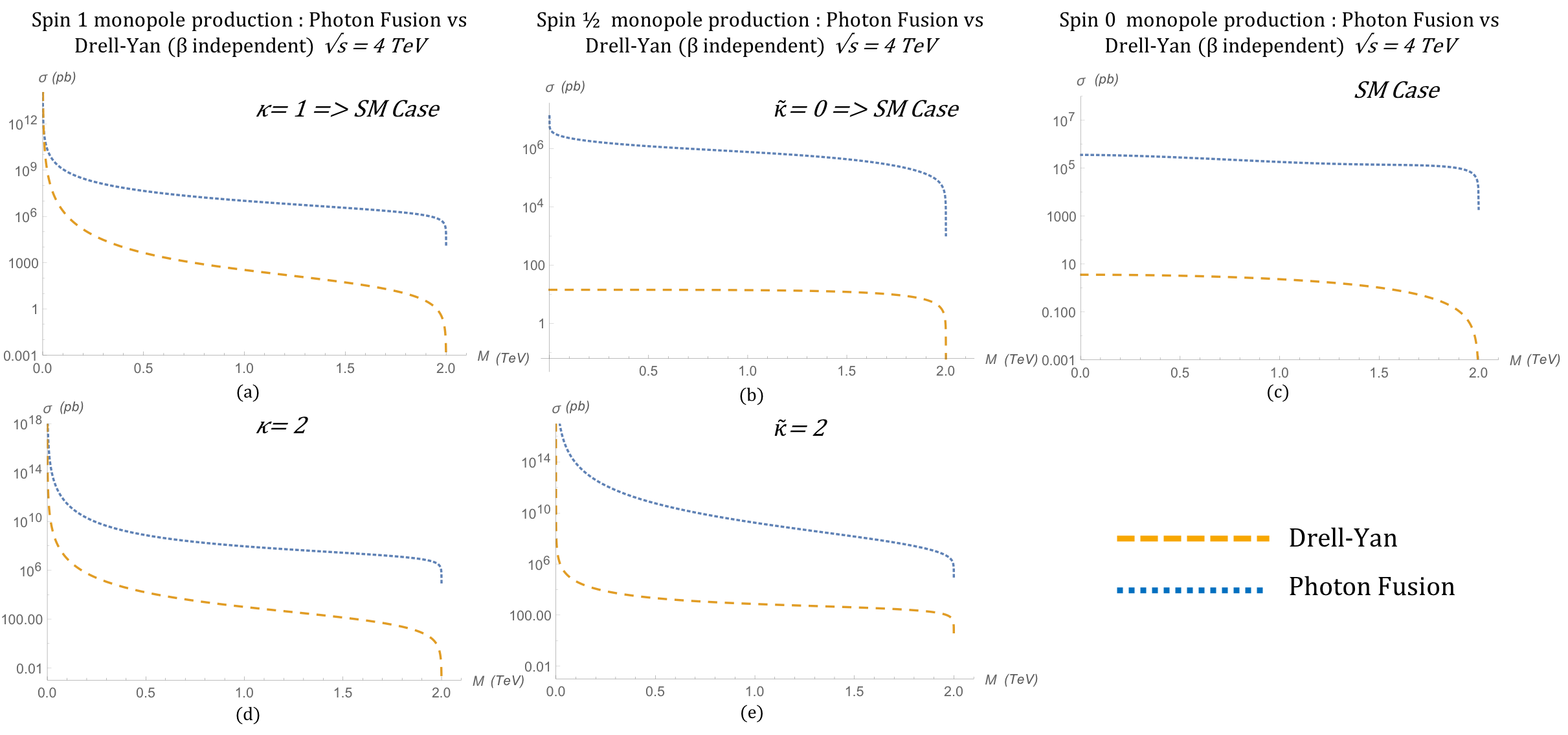}\centering
\caption{Monopole-pair cross sections for PF and DY processes versus mass at $\sqrt{s_{qq/\gamma\gamma}}=4$ TeV for $\beta$-independent coupling. (a) spin-1 in SM-like case ($\kappa=1$); (b) spin-\half in SM-like case ($\tilde{\kappa}=0$); (c) spin-0, without magnetic moment; (d) spin-1 with $\kappa=2$; (e) spin-\half with $\tilde{\kappa}=2$~\cite{Baines:2018ltl}. }\label{PFvsDY}
\end{figure}

The dominant production process is PF by a large margin at $\sqrt{s_{qq/\gamma\gamma}}=4$~TeV, as seen in fig.~\ref{PFvsDY} for all spin models. In particular, graphs~\ref{PFvsDY}(a) and~\ref{PFvsDY}(b) show the SM-like cases for which $\kappa=1$ and $\tilde{\kappa}=0$ represent the $S=1$ and $S=\half$ SM-like cases, respectively; graph~\ref{PFvsDY}(c) shows the spin-0 monopole case. Graphs~\ref{PFvsDY}(d) and~\ref{PFvsDY}(e) demonstrate that the trend is maintained for all values of $\kappa$ and $\tilde{\kappa}$ by choosing the non-distinctive value of two. The cross section for monopole production increases with the spin of the monopole most of the mass range, if the SM-like cases for the magnetic-moment parameters are chosen, supporting the findings of ref.~\cite{dw}.

\subsection{Perturbatively consistent  limiting case of large $\kappa$ and small $\beta$} 
\label{kappaThe}

The non-perturbative nature of the large magnetic Dirac charge invalidates any perturbative treatment of the pertinent cross sections based on Feynman-diagrams. This situation may be resolved if thermal production in heavy-ion collisions is considered~\cite{affleck,arttu}. Another approach involves specific limits of the parameters $\kappa$ and $\beta$ of the effective models of vector and spinor monopoles, for a \emph{velocity-dependent magnetic charge}~\cite{Baines:2018ltl}. In this limit, the perturbative truncation of the production processes becomes meaningful provided the monopoles are slowly moving, that is $\beta \ll 1$.  In terms of the centre-of-mass energy $\sqrt{s_{\gamma\gamma/qq}}$, such a condition on $\beta$ implies, on account of~\eqref{btos}, that the monopole mass is around $2M \simeq \sqrt{s_{\gamma\gamma/qq}} + \mathcal O (\beta^2)$. We remark that, in hadron-collider production of monopole--antimonopole pairs, $s_{\gamma\gamma/qq}$ is not definite but follows a distribution defined by the parton/photon distribution function (PDF) for the DY or PF processes.  

In the absence of a magnetic-moment parameter, $\kappa$, or for the unitary value $\kappa=1$ in the case of spin-1 monopoles, the condition $\beta \ll 1$ would lead to strong suppression of the pertinent cross sections beyond the current experimental sensitivities, thereby rendering the limit $\beta \to 0$ experimentally irrelevant for placing bounds on monopole masses or magnetic charges.
Indeed, the various total cross sections discussed so far behave as follows, when $\beta \to 0$:
\begin{alignat}{5}\label{reproRusakovich1b}
 \sigma_{\gamma\gamma\rightarrow M\overline{M}}^{S=1, \kappa=1} \quad  & \stackrel{\beta \to 0}{\simeq} && \quad 
\frac{19\, g^4}{8\pi\, s}\, \beta^5 \quad && \xrightarrow{\beta \to 0} && \quad 0 \qquad \text{(spin-1~PF)},  \nonumber \\
 \sigma^{S=1, \kappa=1}_{q\overline{q}\rightarrow M\overline{M}}   \quad  & \stackrel{\beta \to 0}{\simeq} && \quad 
\frac{135 \,  s \, \alpha _e \, g^2}{1728\,  M^4}\beta^5 \quad && \xrightarrow{\beta \to 0} && \quad 0 \qquad \text{(spin-1~DY)}, \nonumber \\
 \sigma_{\gamma\gamma\rightarrow M\overline{M}}^{S=\frac{1}{2}, \kappa=0} \quad  & \stackrel{\beta \to 0}{\simeq} && \quad  \frac{g^4}{4\pi\, s} \, \beta^5 
\quad && \xrightarrow{\beta \to 0} && \quad 0 \qquad \text{(spin-\half~PF)}, \nonumber \\
  \sigma^{S=\frac{1}{2}, \kappa=0}_{q\overline{q}\rightarrow M\overline{M}}   \quad  & \stackrel{\beta \to 0}{\simeq} && \quad  \frac{5\, \alpha_e \, g^2}{18\,  \, s}\, \beta^3 \quad && \xrightarrow{\beta \to 0} && \quad 0 \qquad \text{(spin-\half~DY)}, \nonumber \\
 \sigma^{S=0}_{\gamma\gamma\rightarrow \overline{M}M} \quad  & \stackrel{\beta \to 0}{\simeq} && \quad \frac{g^4}{4\pi\, s}\, \beta^5 \, && \xrightarrow{\beta \to 0} && \quad  0  \qquad \text{(spin-0~PF)},
\nonumber \\
  \sigma_{q\overline{q}\rightarrow M\overline{M}}^{S=0}  \quad & \stackrel{\beta \to 0}{\simeq} && \quad 
  \frac{5\,\alpha_{e}\, g^2}{108 \, s}\, \beta^5 \quad && \xrightarrow{\beta \to 0} && \quad  0  \qquad \text{(spin-0~DY)}. 
\end{alignat}

However, in the case of non-trivial and \emph{large} (dimensionless) magnetic-moment-related parameters $\kappa, \tilde \kappa$, relevant for the cases of vector and spinor monopoles, the situation changes drastically~\cite{Baines:2018ltl} if one  considers the limits 
\begin{equation}\label{limit}
\kappa \gg 1, \quad \tilde \kappa \gg 1, \quad \beta \ll 1, 
\end{equation}
with $\tilde \kappa$ defined in \eqref{ktilde},
but in such a way that the strength of the derivative magnetic-moment couplings for spin-\half monopoles is perturbatively small. Since the magnitude of the monopole momentum is proportional to $M \beta$, one expects that the condition of a perturbatively small derivative coupling for magnetic moment is guaranteed if, by order of magnitude, one has:
\begin{equation}\label{kbg}
g  \kappa^\prime  \beta^2  < 1  ~, 
\end{equation}
where $\kappa^\prime = \tilde\kappa$ for spin-\half monopole and $\kappa^\prime = \kappa$ for spin-1 monopole. 

For the spin-\half monopole, one observes that in the limit \eqref{limit} and respecting \eqref{kbg}, upon requiring the absence of infrared divergences in the cross sections as $\beta \to 0$, and postulating that:
\begin{equation}\label{ktildelim}
(\tilde \kappa \beta g)^4  \beta \xlongequal{\stackrel{\beta \to 0}{\kappa \to \infty}}  |c^\prime_1|, \qquad c^\prime_1 = \text{finite constant},
\end{equation}
so that \eqref{kbg} is trivially satisfied, since $\tilde \kappa g \beta^2   = |c^\prime_1|^{\frac{1}{4}}  \beta^{\frac{3}{4}} \, \xrightarrow{\stackrel{\beta \to 0}{\kappa \to \infty}} 0$,
the dominant contributions to the PF and DY total cross sections  are   
given by 
\begin{equation}\label{totsec12lim}
\begin{split}
 \sigma^{S=\frac{1}{2}}_{\gamma\gamma\rightarrow M\overline{M}} &\sim\pi  \alpha_g^2(\beta) \beta \kappa^4 s = \frac{(\tilde \kappa \, g \, \beta)^4  \, \beta }{16 \pi M^4} s \xlongequal{\stackrel{{\beta \to 0}}{\tilde \kappa \to \infty}} \text{finite},
\end{split}
\end{equation}
and
\begin{equation}\label{totsecdy12lim}
\sigma^{S=\frac{1}{2}}_{q\overline{q}\rightarrow M\overline{M}}\sim\pi\alpha_e \alpha_g(\beta) \, \frac{10\beta\kappa^2}{9} = \frac{5 \alpha_e}{18M^2}  \, 
(\tilde \kappa \beta  g )^2 \beta   \xrightarrow{\stackrel{\beta \to 0}{\kappa \to \infty}} \, 0,
\end{equation}
respectively. Hence, for slowly-moving spinor-monopoles, with velocity-dependent magnetic charge, and large magnetic moment parameters, it is the PF cross section which is the dominant one of relevance to collider experiments~\cite{Baines:2018ltl}. 

Similar results characterise the spin-1 monopole. Indeed, in the limit \eqref{limit}, \eqref{kbg}, the dominant contributions to the total cross sections for the PF  and DY processes are such that
\begin{equation}\label{totxsec1limit}
\begin{split}
 \sigma^{S=1}_{\gamma\gamma\rightarrow M\overline{M}}&\sim\pi  \alpha_g^2 \, \frac{29 \beta^5 \, \kappa^4}{4 s} = \frac{29}{64\pi s} \beta\,  \big(\kappa \beta g \big)^4 ,
\end{split}
\end{equation}
and 
\begin{equation}\label{dytotsec1lim}
\sigma^{S=1}_{q\overline{q}\rightarrow M\overline{M}}=\alpha _e \alpha _g(\beta)\pi \frac{40\beta^3\kappa^2}{27s} 
= \alpha_e \, \frac{10}{27s} \big(\kappa \beta g \big)^2  \beta^3,
\end{equation}
respectively. We can see that, by requiring the \emph{absence of infrared} ($\beta \to 0$) \emph{divergences} in the total cross sections, one may consistently arrange that
the PF cross section \eqref{totxsec1limit} acquires a non-zero (finite) value as $\beta \to 0$, whilst the DY cross section \eqref{dytotsec1lim} vanishes in this limit:
\begin{eqnarray}\label{limits}
(\kappa \beta  g)^4  \beta & \xlongequal{\stackrel{\beta \to 0}{\kappa \to \infty}} &  |c_1|, \qquad c_1 = \text{finite constant}, \nonumber \\
\sigma^{S=1}_{\gamma\gamma\rightarrow M\overline{M}} 
& \xlongequal{\stackrel{\beta \to 0}{\kappa \to \infty}} &
\frac{29\, c_1}{64\, \pi \, s} ,
\nonumber \\
\sigma^{S=1}_{q\overline{q}\rightarrow M\overline{M}} 
& \xlongequal{\stackrel{\beta \to 0}{\kappa \to \infty}}&
\alpha_e \, \frac{10\, \sqrt{|c_1|}}{27\,s} \, \beta ^{\frac{5}{2}} \quad \xrightarrow{\stackrel{\beta \to 0}{\kappa \to \infty}}  \quad 0.
\end{eqnarray}
In such a limit, the quantity $\kappa g \beta^2  = |c_1|^{\frac{1}{4}} \, \beta^{\frac{3}{4}} \, \xrightarrow{\stackrel{\beta \to 0}{\kappa \to \infty}} \, 0$, so \eqref{kbg} is trivially satisfied, and thus  the perturbative nature of the magnetic moment coupling is guaranteed. 
Hence in this limiting case of velocity-dependent magnetic charge, large magnetic moment couplings and slowly moving vector monopoles, again the PF cross section is the dominant one relevant to searches in current colliders and can be relatively large (depending on the value of the phenomenological parameter $c_1$)~\cite{Baines:2018ltl}. 

\section{\MAD implementation and LHC phenomenology\label{sec:mad}}

The \MAD generator~\cite{MG} is used to simulate monopoles by developing a dedicated Universal FeynRules Output (UFO) model~\cite{ufo}, as explained in detail in ref.~\cite{Baines:2018ltl}. This includes both the $\beta$-dependent and -independent photon--monopole coupling. Three different spin cases have been considered:  spins 0, \half and 1. To generate a UFO model from the Lagrangian, \Feyn~\cite{feyn2}, an interface to \Math, was utilised. The velocity $\beta$, defined in \eqref{btos}, is defined as a form factor in the generated UFO models. For scalar and vector monopoles, the inclusion in the simulation of the four-particle vertex shown in fig.~\ref{fig:seagull} in addition to the $u$- and $t$-channel, shown in  figs.~\ref{fig:uchan} and~\ref{fig:tchan} respectively, led to the necessary use of UFO model written as a \pyt object. The implementation of the four-vertex diagram proved to be non-trivial due to the $g^2$ coupling.

Apart from the photon fusion production mechanism, the UFO models were also tested for the DY production mechanism as well. The DY process for monopoles was already implemented in \MAD using a \fort code setup for spin-0 and spin-\half monopoles, utilised by ATLAS~\cite{atlasmono1,atlasmono2} and MoEDAL~\cite{moedalplb,moedal}. These \fort setups were rewritten in the context of this work as UFO models, following their PF counterparts, and they were extended to include the spin-1 case. The new implementation for PF and DY was used in the most recent MoEDAL analysis~\cite{Acharya:2019vtb}. After validating the UFO models against their \fort implementations for scalar and fermionic monopoles, all spin cases were compared by the theoretical predictions of section~\ref{sec:spin} and of ref.~\cite{Baines:2018ltl}. 

\subsection{Comparison between the photon fusion and the Drell--Yan production mechanisms}\label{sec:kinematics}

Apart from the total cross sections, it is important to study the angular distributions of the generated monopoles. This is of great interest to the interpretation of the monopole searches in collider experiments, given that the geometrical acceptance and efficiency of the detectors is not uniform as a function of the solid angle around the interaction point. 
Here the kinematic distributions are compared between the photon-fusion ($\gamma\gamma$) and the Drell--Yan mechanisms. To this end, the $\beta$-dependent UFO monopole model was used in \MAD. Monopole--antimonopole pair events have been generated for $pp$ collisions at $\sqrt{s}=13$~TeV, i.e.\ for the LHC Run-2 operating energy. The PDF was set to \texttt{NNPDF23}~\cite{nnpdf} at LO for the Drell--Yan and \texttt{LUXqed}~\cite{lux} for the photon-fusion mechanism. The monopole magnetic charge is set to 1~\gd, yet the kinematic spectra are insensitive to this parameter.

\begin{figure}[ht!]
\justify
\includegraphics[width=0.33\textwidth]{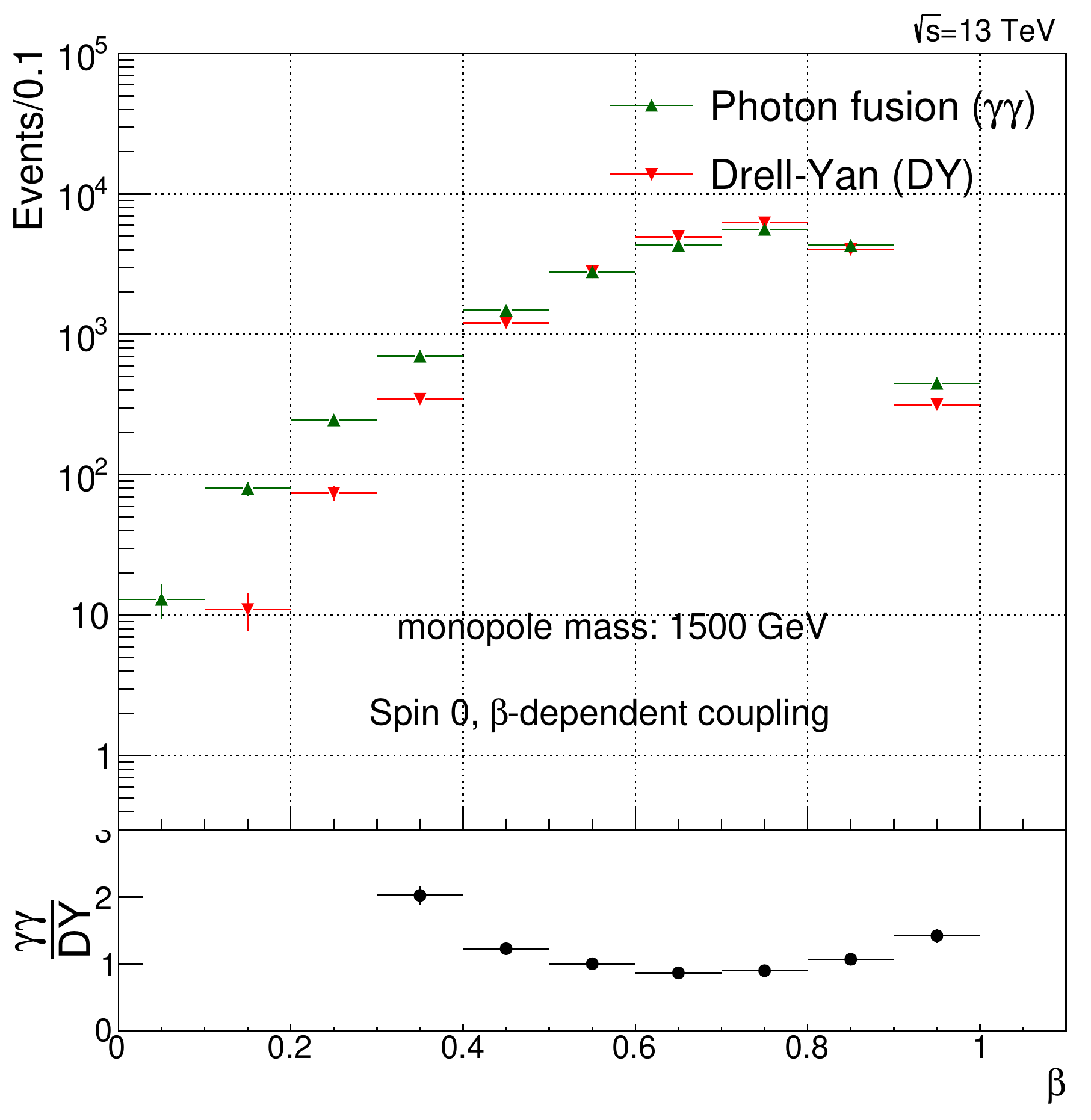}\hfill
\includegraphics[width=0.33\textwidth]{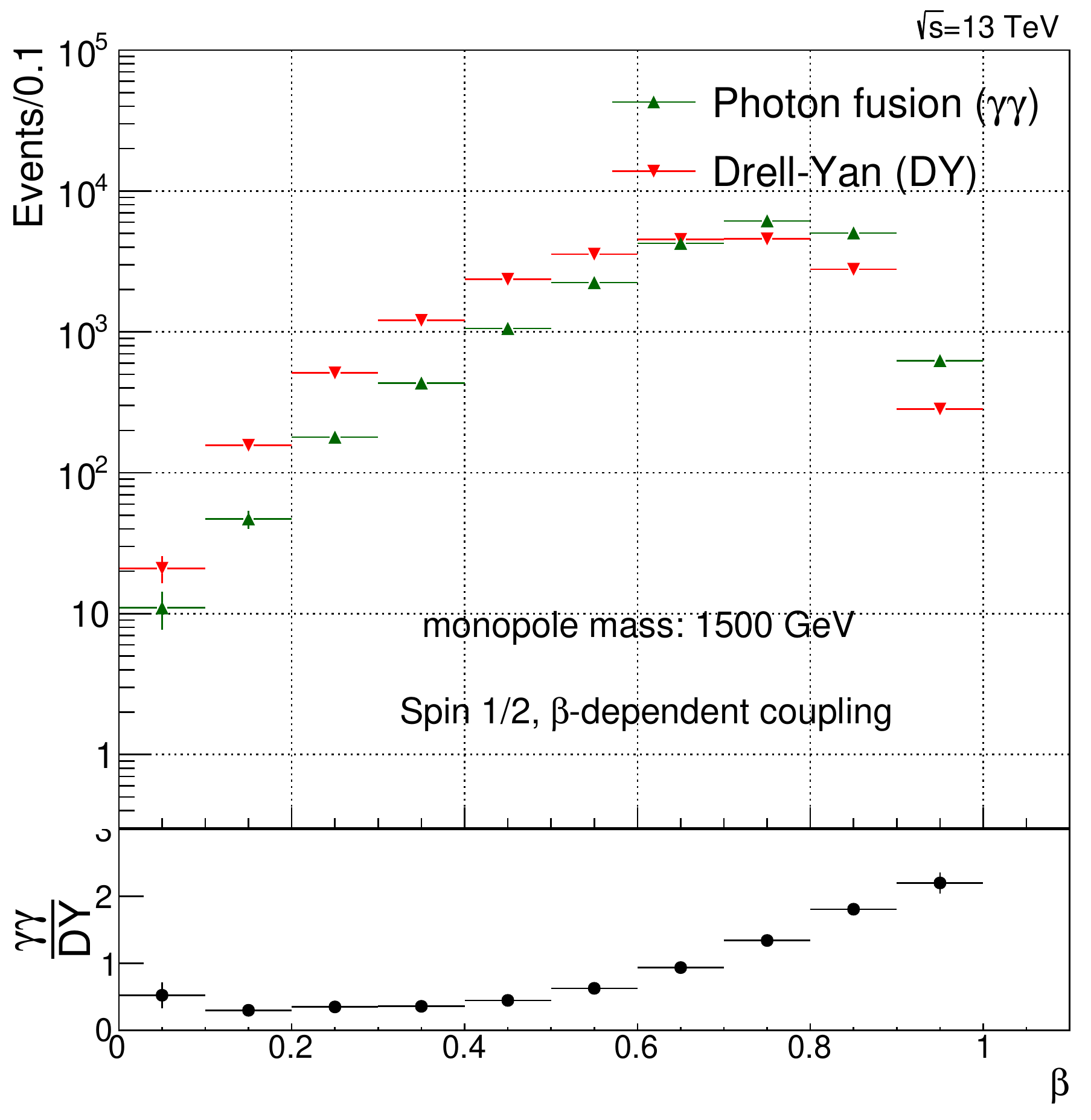}\hfill
\includegraphics[width=0.33\textwidth]{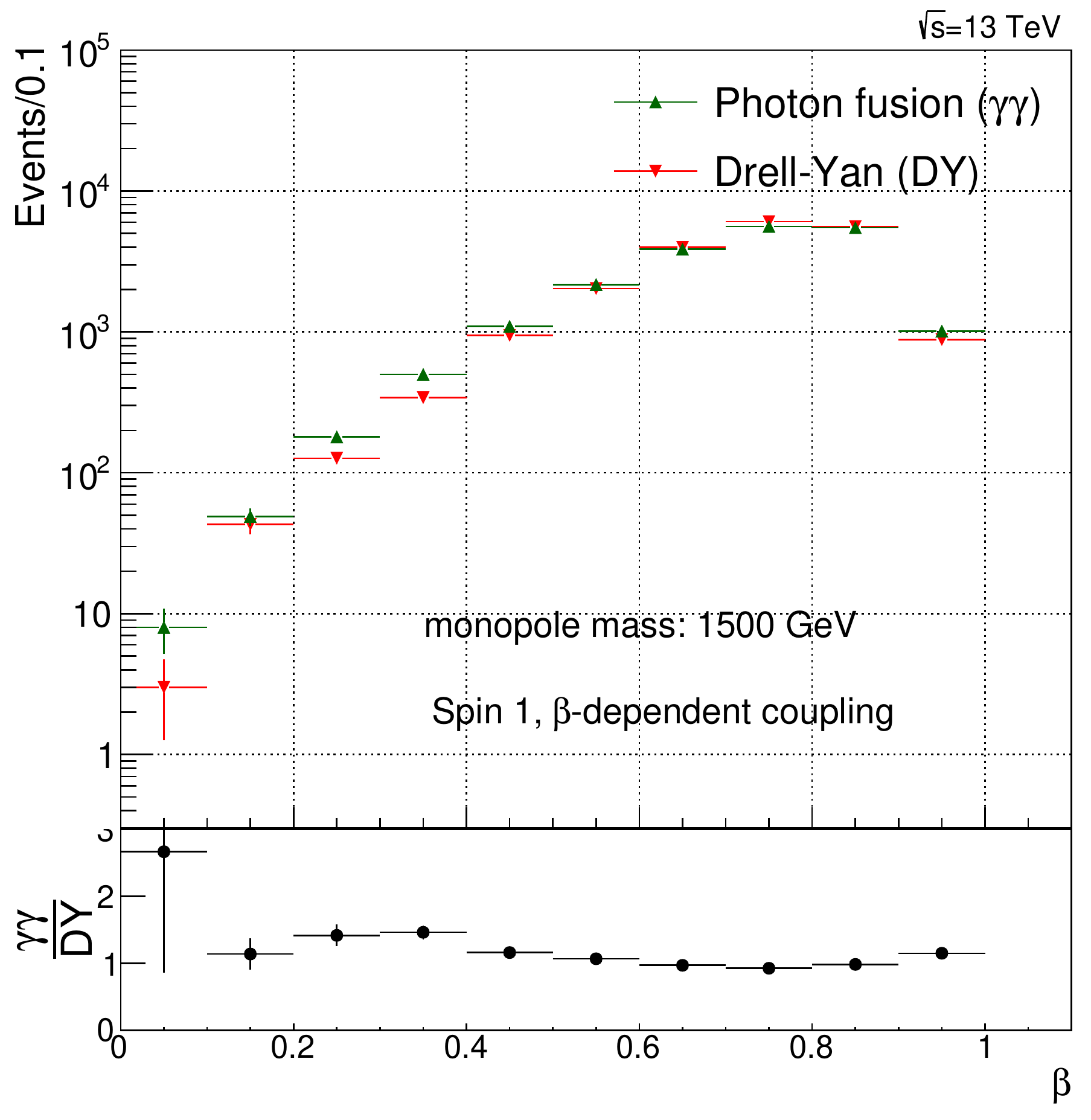}
\caption{The monopole velocity $\beta$ distributions of spin-0 (left), spin-\half (centre) and spin-1 monopoles (right) for both the photon-fusion and the Drell--Yan mechanisms at $\sqrt{s}=13$~TeV~\cite{Baines:2018ltl}.   }\label{fig:beta}
\end{figure}

The distributions of the monopole velocity, a crucial parameter for the detection of monopoles in experiments such as MoEDAL~\cite{moedal-review}, are depicted in fig.~\ref{fig:beta}. The velocity $\beta$, which is calculated in the centre-of-mass frame of the colliding protons, largely depends on the PDF of the photon ($q$/$\bar{q}$) in the proton for the photon-fusion (Drell--Yan) process. For scalar monopoles, fig.~\ref{fig:beta} (left) shows that slower-moving monopoles are expected for PF than DY, an observation favourable for the discovery potential of MoEDAL NTDs, the latter being sensitive to low-$\beta$ monopoles. The comparison is reversed for fermionic monopoles, where PF yields faster monopoles than DY (fig.~\ref{fig:beta} (centre)). Last, as deduced from fig.~\ref{fig:beta} (right), the $\beta$ distributions for PF and DY are very similar. 

\begin{figure}[ht!]
\justify
\includegraphics[width=0.33\textwidth]{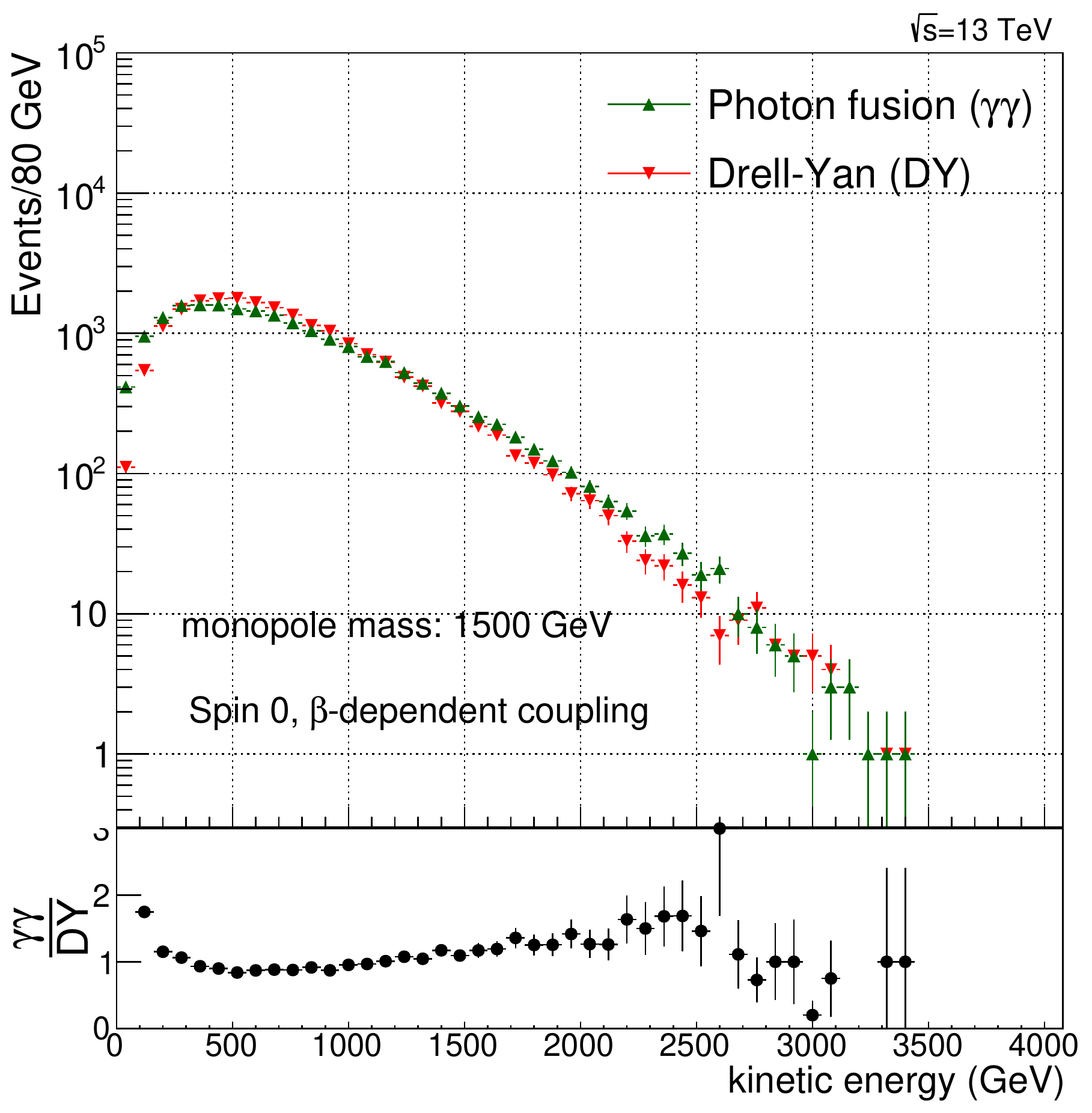}\hfill
\includegraphics[width=0.33\textwidth]{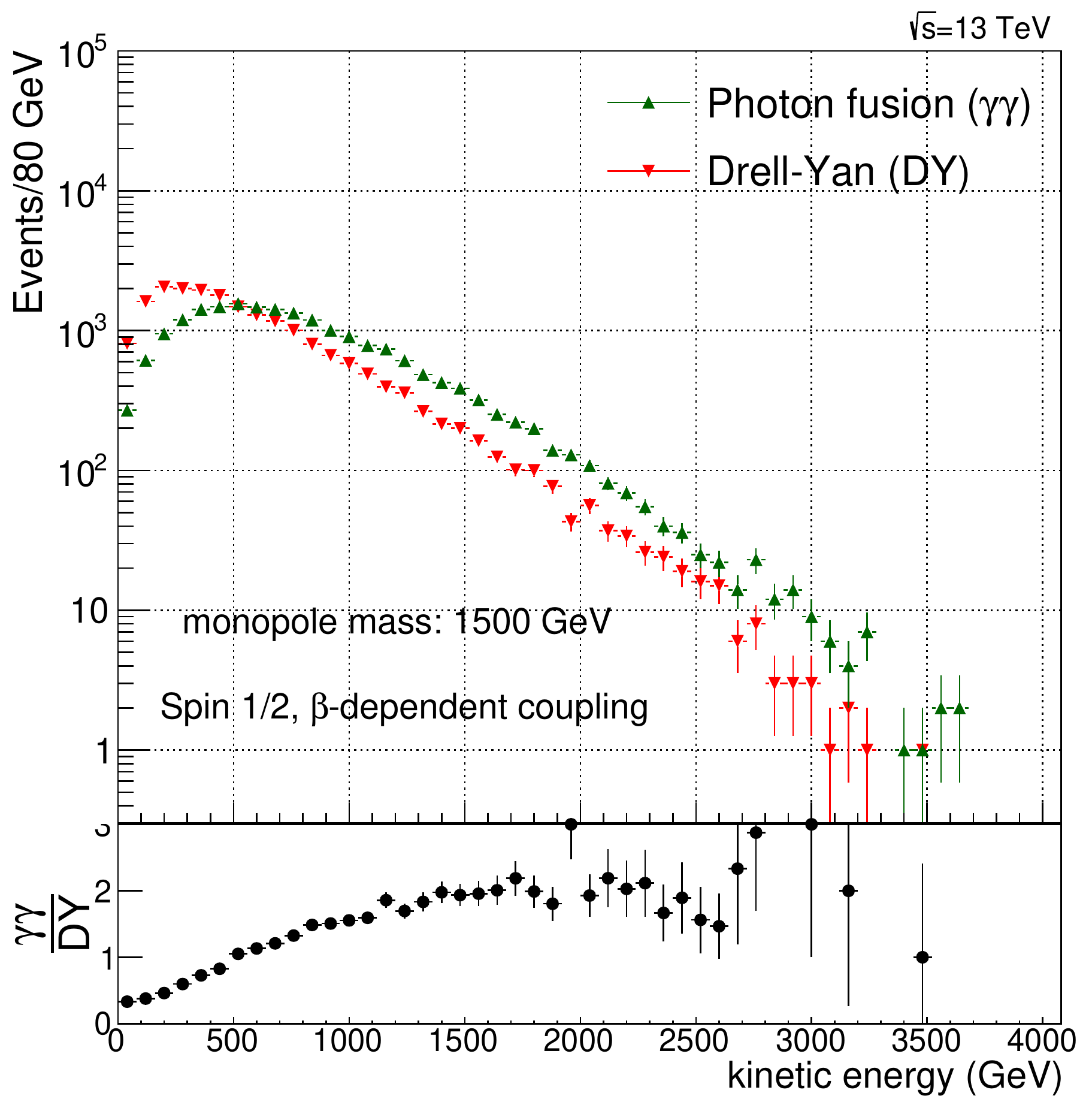}\hfill
\includegraphics[width=0.33\textwidth]{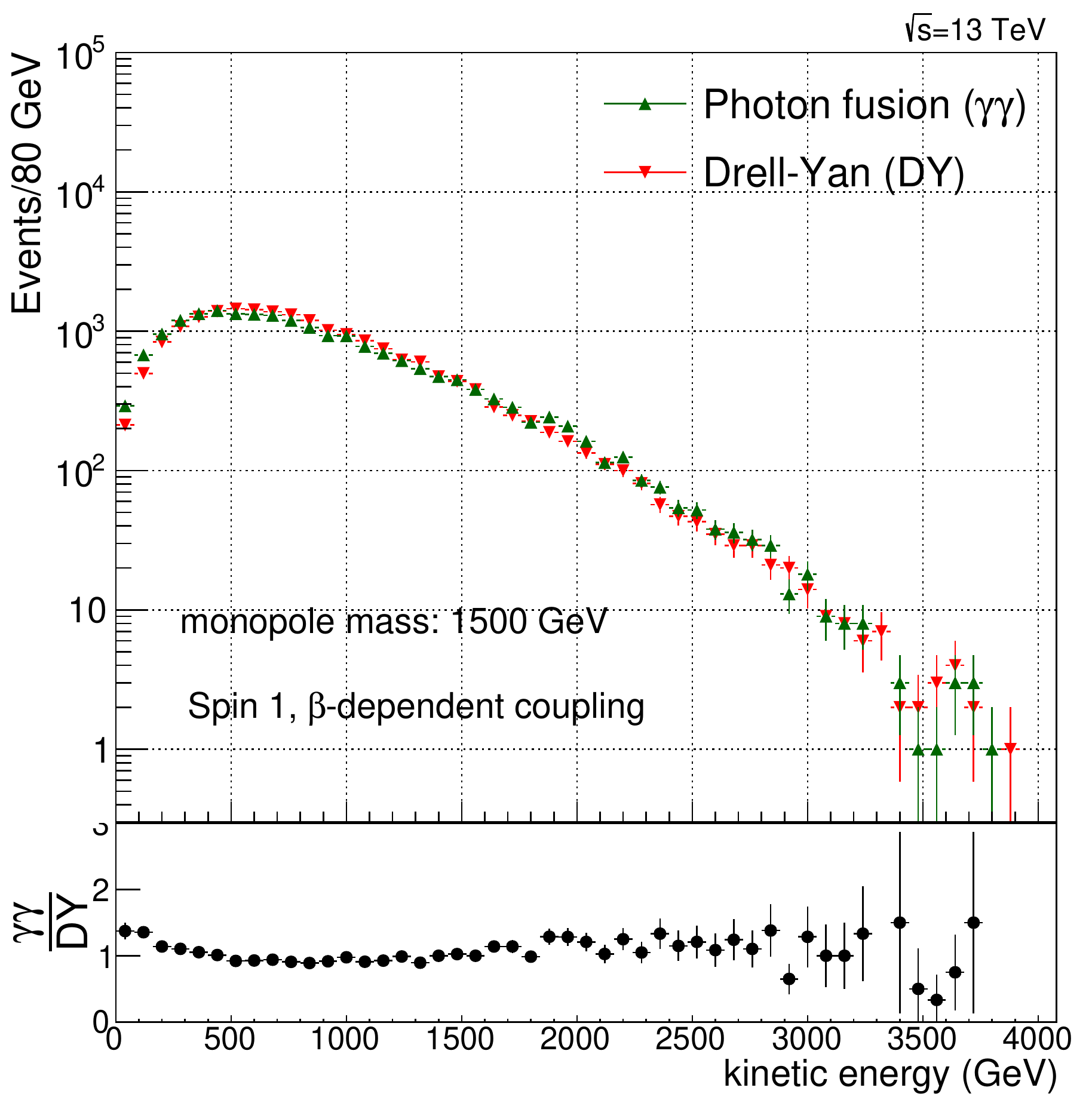}
\caption{The kinetic energy distributions for spin-0 (left), spin-\half (centre) and spin-1 monopoles (right) for the photon-fusion and the Drell--Yan mechanisms at $\sqrt{s}=13$~TeV and for $\tilde{\kappa}=0$~\cite{Baines:2018ltl}.  }\label{fig:ke}
\end{figure}

The kinetic-energy spectrum, shown in fig.~\ref{fig:ke}, is slightly softer for PF than DY for scalar (left panel) and vector (right panel) monopoles, whereas it is significantly harder for fermions (central panel). This difference may be also due to the four-vertex diagram included in the bosonic monopole case. This observation is in agreement with the one made for $\beta$ previously.  We have also compared \MAD predictions for kinetic-energy and \pt distributions between with- and without-PDF cases, the latter also against analytical calculations (cf.\ section~\ref{sec:spin}) across different spins and production mechanisms. As expected, some features seen in the direct $\gamma\gamma$ or $q\bar{q}$ production are attenuated in the $pp$ production due to the sampling of different $\beta$ values in the latter as opposed to the fixed value in the former. 

\begin{figure}[ht!]
\justify
\includegraphics[width=0.33\textwidth]{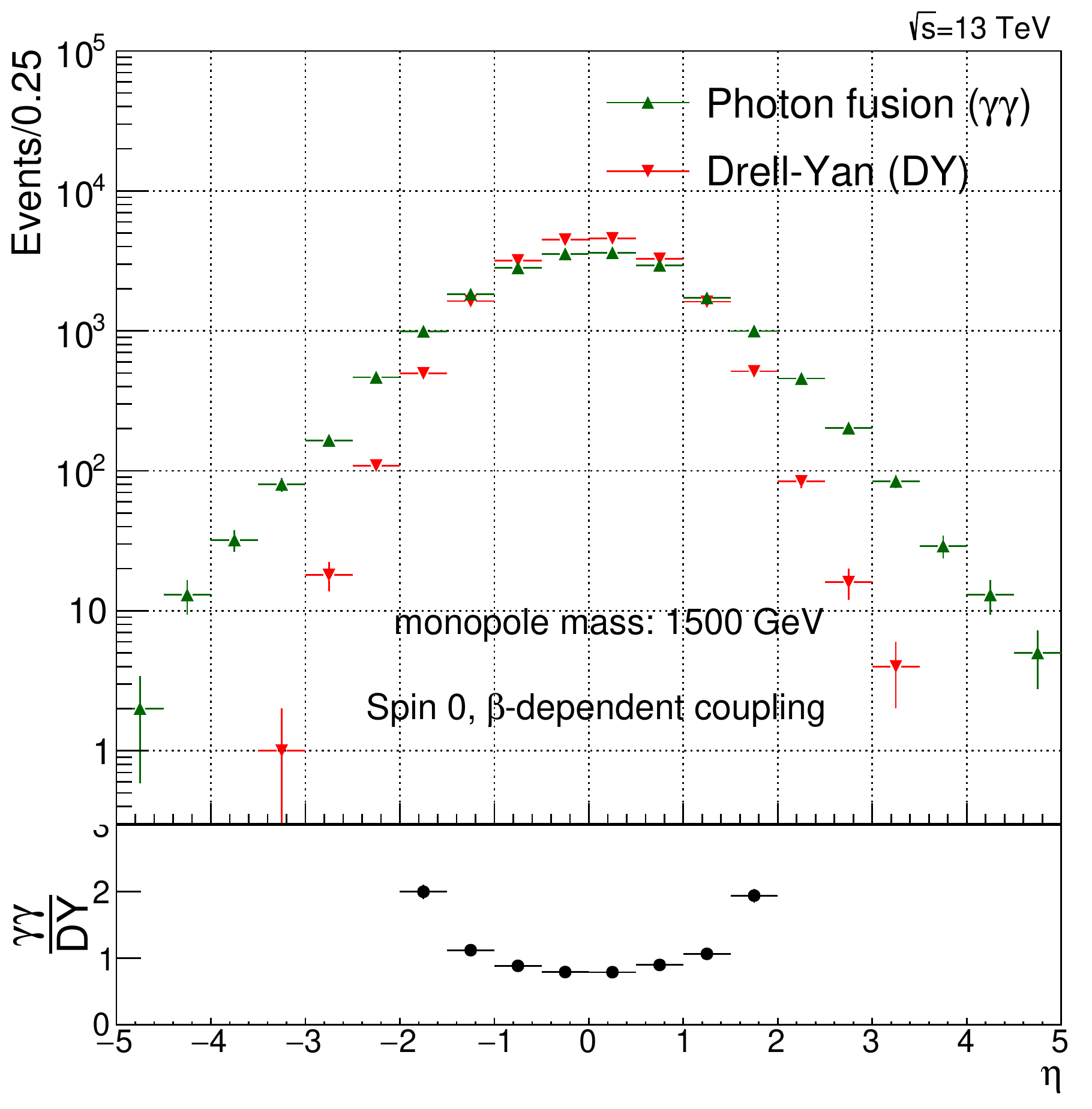}\hfill
\includegraphics[width=0.33\textwidth]{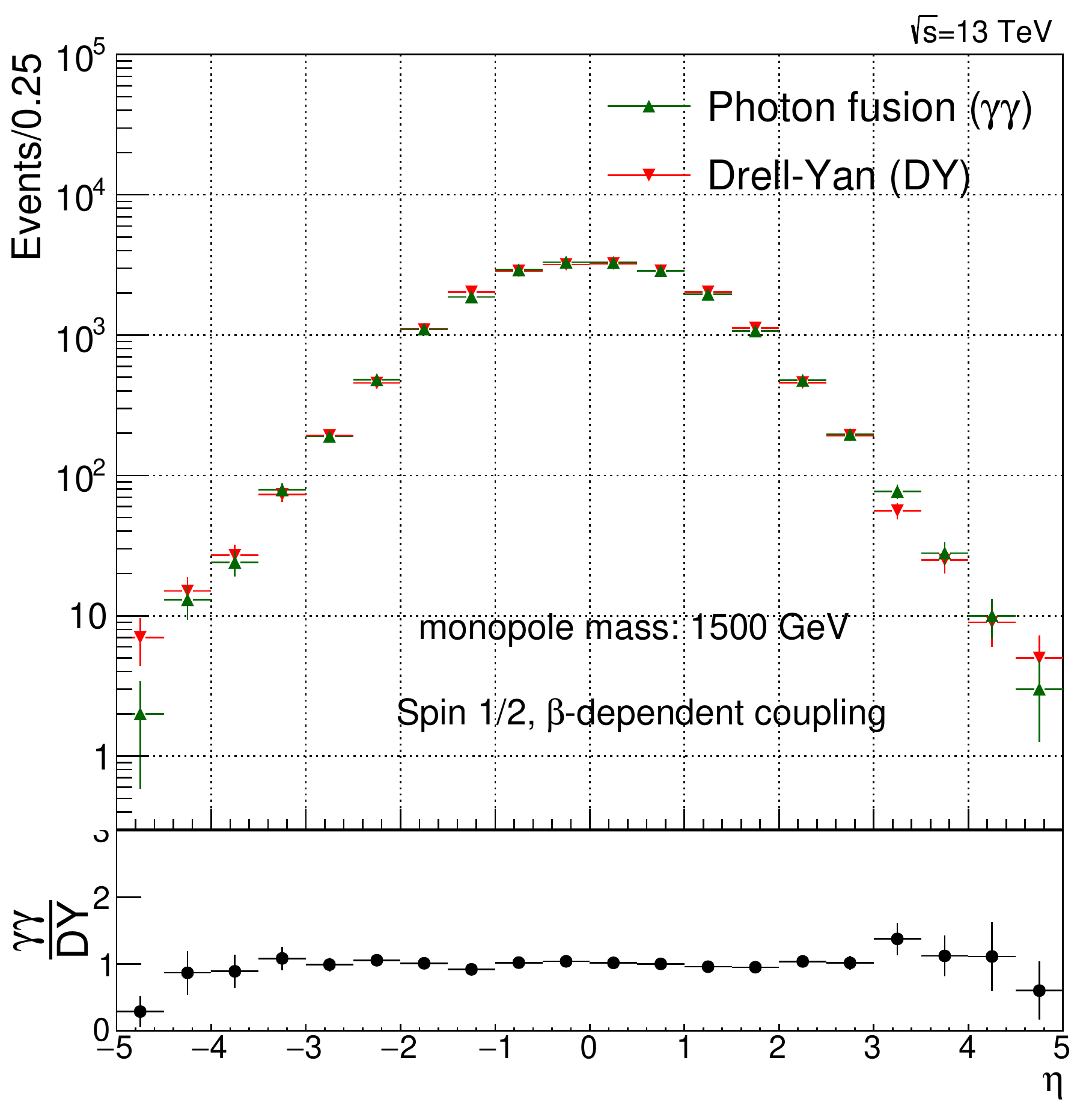}\hfill
\includegraphics[width=0.33\textwidth]{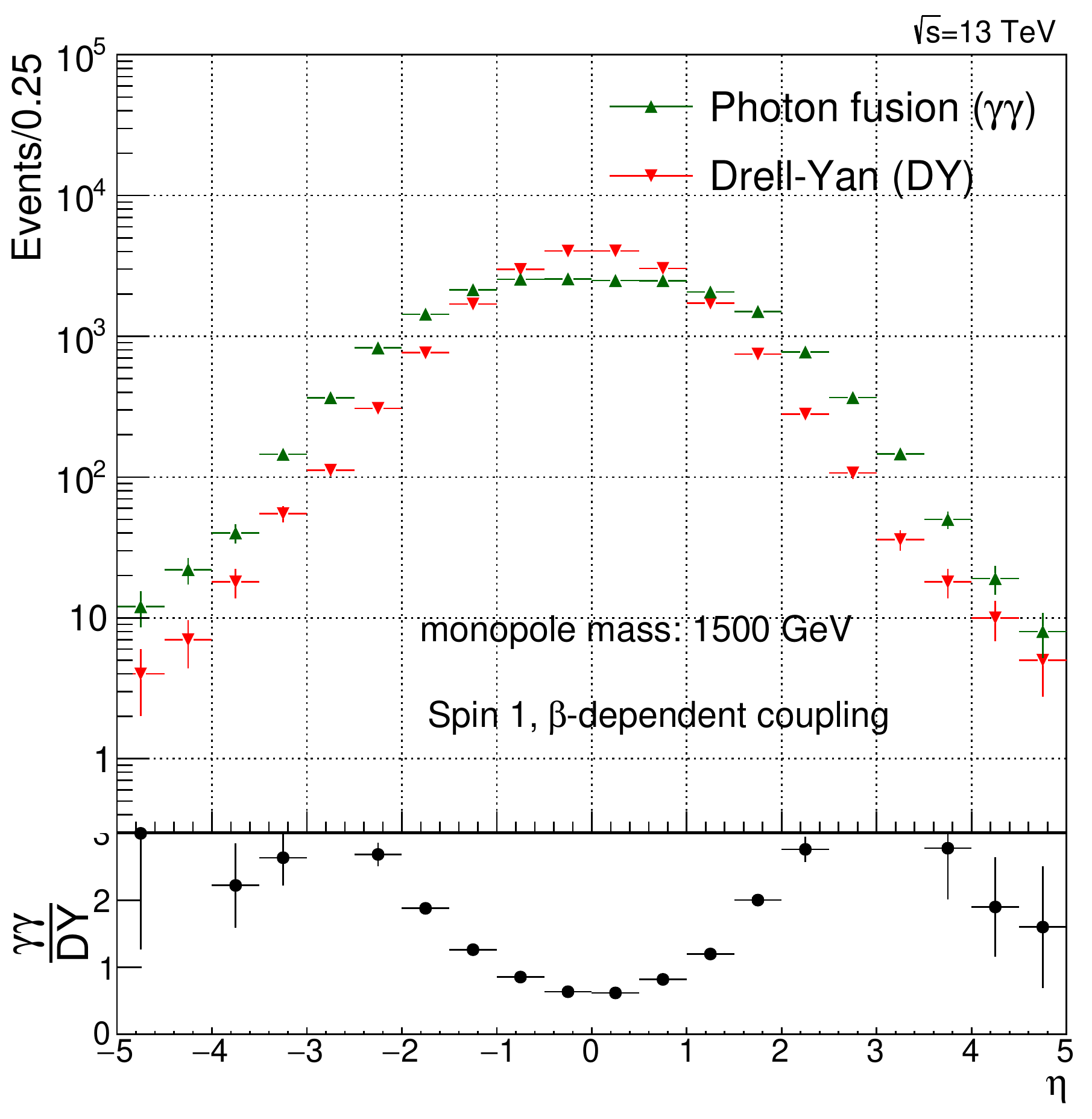}
\caption{The pseudorapidity $\eta$ distributions for spin-0 (left), spin-\half (centre) and spin-1 monopoles (right) for the photon-fusion and the Drell--Yan mechanisms at $\sqrt{s}=13$~TeV and for $\kappa=1$~\cite{Baines:2018ltl}.  }\label{fig:eta}
\end{figure}

As far as the pseudorapidity is concerned, its distributions are shown in fig.~\ref{fig:eta}. The spin-0 (left panel) and spin-1 (right panel) cases yield a more central production for DY than PF, whilst for spin-\half (central panel) the spectra are practically the same, although the one for PF is slightly more central. Again this behaviour of bosonic versus fermionic monopoles may be attributed to the (additional) four-vertex diagram for the bosons. 

\begin{figure}[ht!]
\justify
\includegraphics[width=0.33\textwidth]{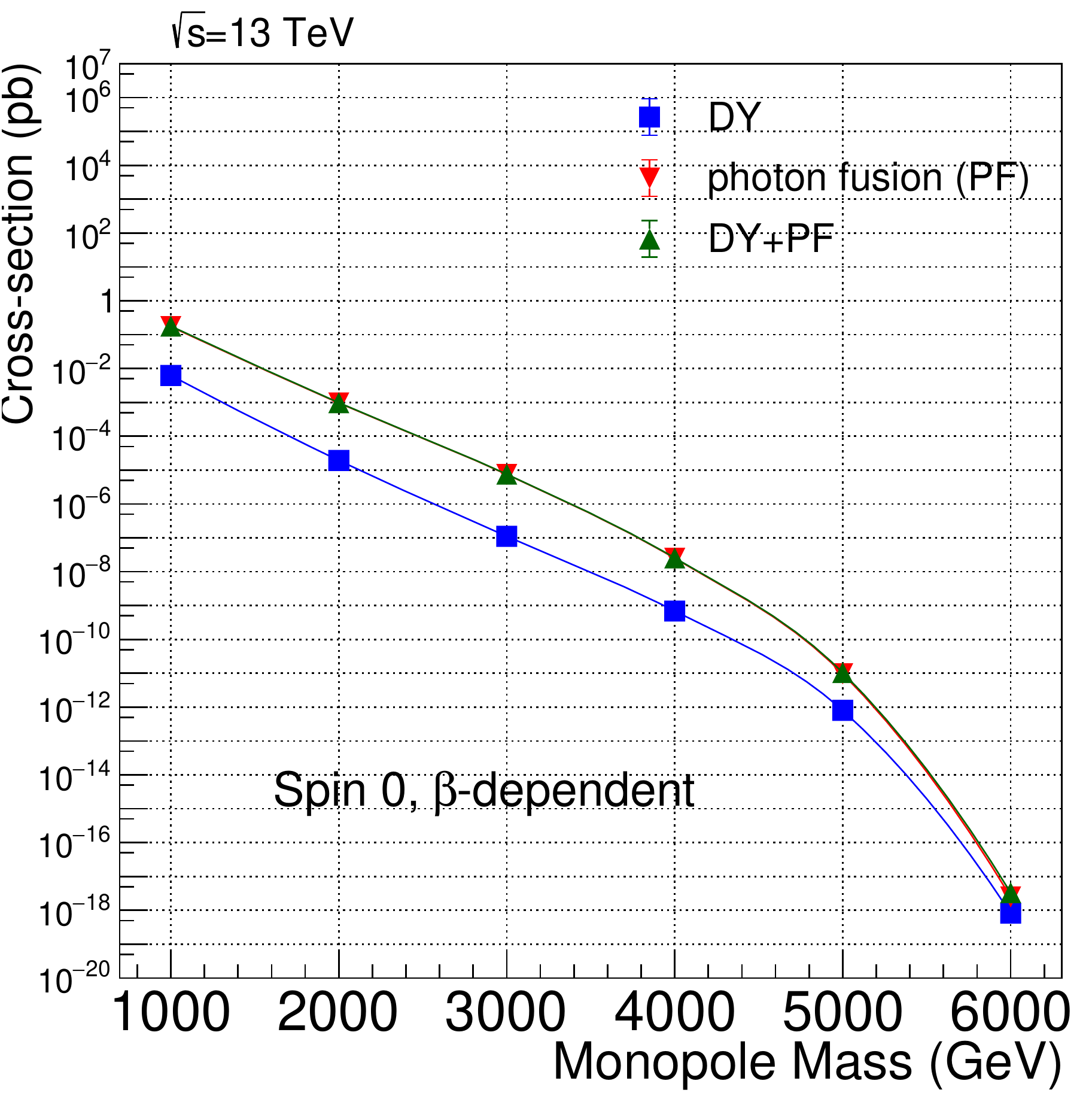}\hfill
\includegraphics[width=0.33\textwidth]{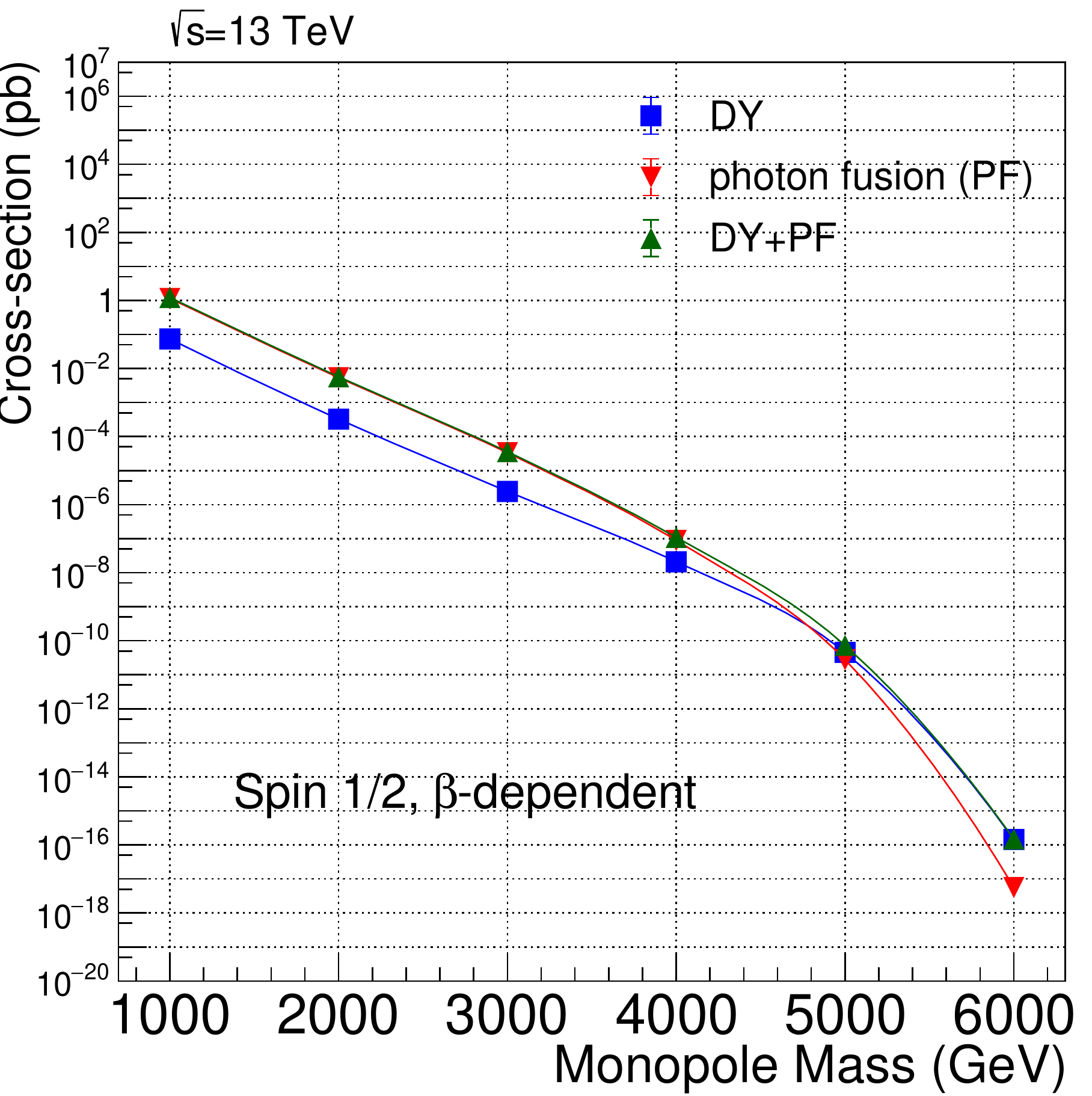}\hfill
\includegraphics[width=0.33\textwidth]{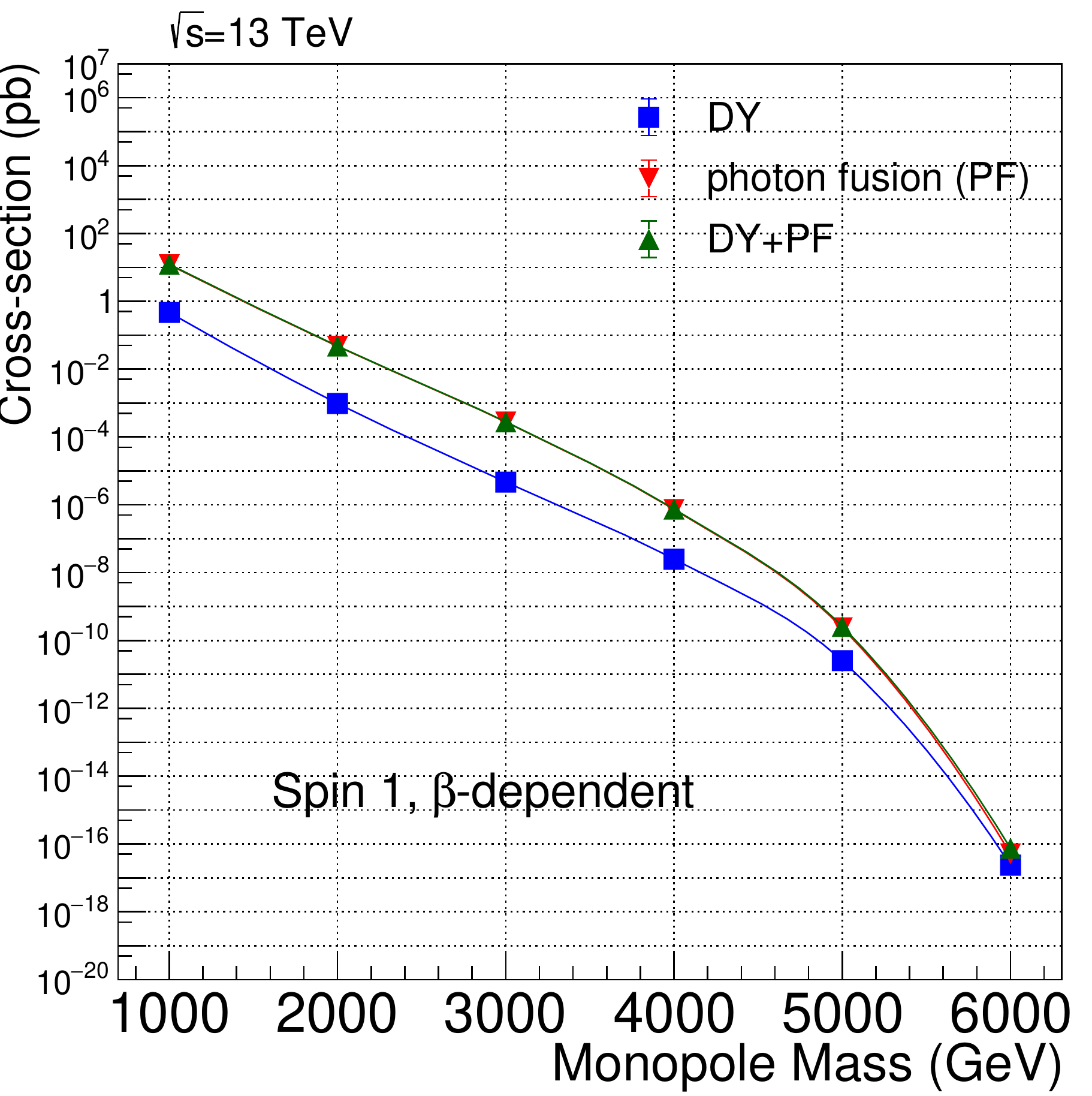}
\caption{Total cross section at $\sqrt{s}=13$~TeV for PF, DY and their sum versus monopole mass.  For $S=\half$ and $S=1$, the SM values $\tilde{\kappa}=0$ and $\kappa=1$, respectively, are drawn, while there is no magnetic moment in the spin-0 case~\cite{Baines:2018ltl}. }\label{fig:xsec-pdf}
\end{figure}

The total cross sections for the various spin cases, assuming SM magnetic-moment values for spin~\half and spin~1 and $\beta$-dependent coupling are drawn in fig.~\ref{fig:xsec-pdf} for photon fusion and Drell--Yan processes, as well as their sum. At tree level there is no interference between the PF and DY diagrams, so the sum of cross sections expresses the sum of the corresponding amplitudes. The PF mechanism is the dominant at the LHC energy of 13~TeV throughout the whole mass range of interest of $1\div 6~{\rm TeV}$ for the bosonic-monopole case. However if the monopole has spin~\half the PF dominates for masses up to $\sim 5~{\rm TeV}$, while DY takes over for $M\gtrsim5~{\rm TeV}$. 

\subsection{Perturbatively consistent limiting case of large $\kappa$ and small $\beta$ for photon fusion}\label{kappaMad}

In section~\ref{kappaThe}, the theoretical calculations show that in the perturbatively consistent limit of large $\kappa$ and small $\beta$, the cross sections are finite for both spin-\half  \eqref{totsec12lim} and spin-1 \eqref{limits} cases. In this section, we focus on this aspect of the photon-fusion production mechanism, since it dominates at LHC energies. We first put to test this theoretical claim utilising the \MAD implementation and later we discuss the kinematic distributions and comment on experimental aspects of a potential perturbatively-consistent search in colliders to follow in this context. 

\begin{figure}[ht!]
\justify
\includegraphics[width=0.33\textwidth]{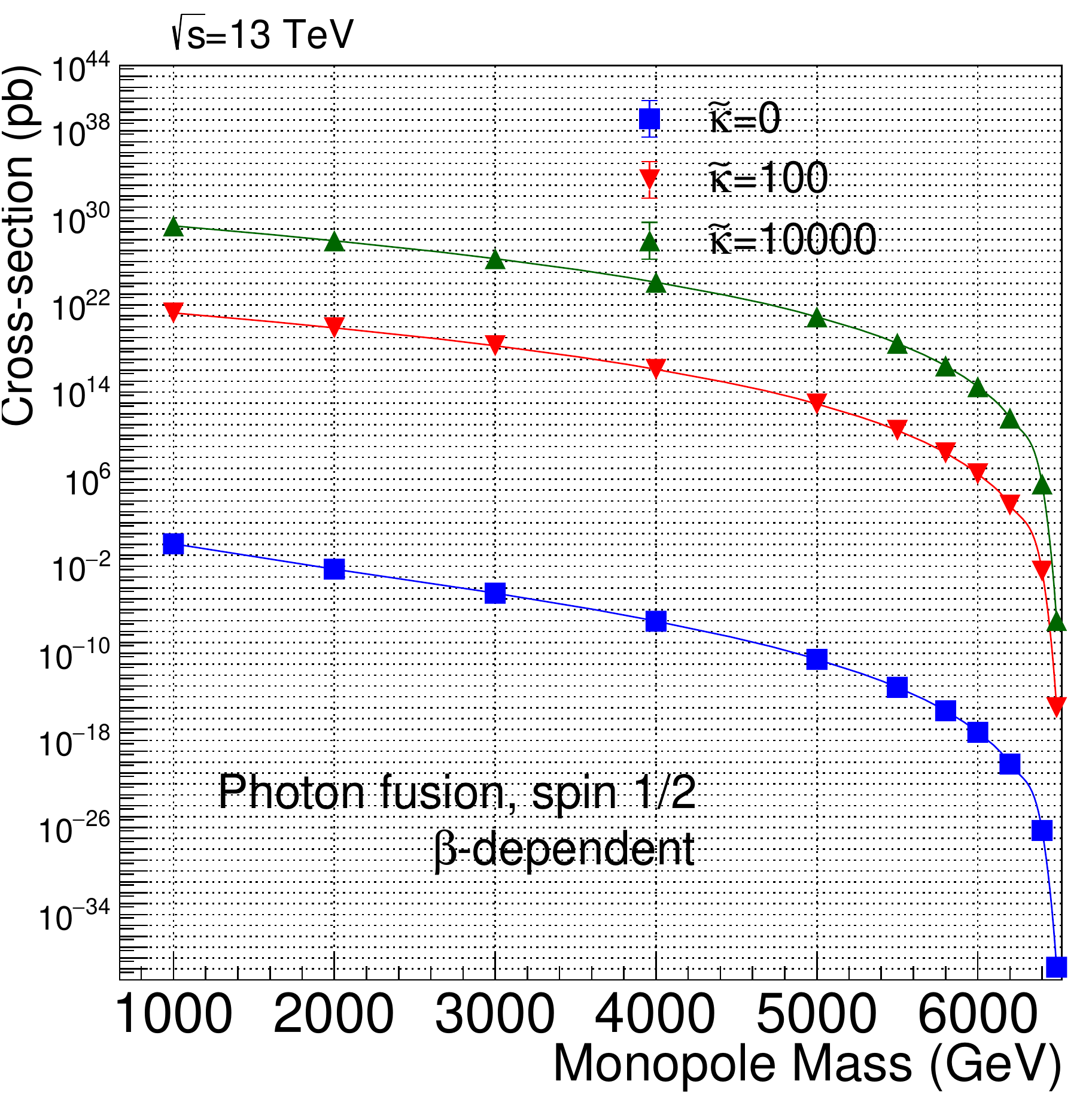}\hfill
\includegraphics[width=0.33\textwidth]{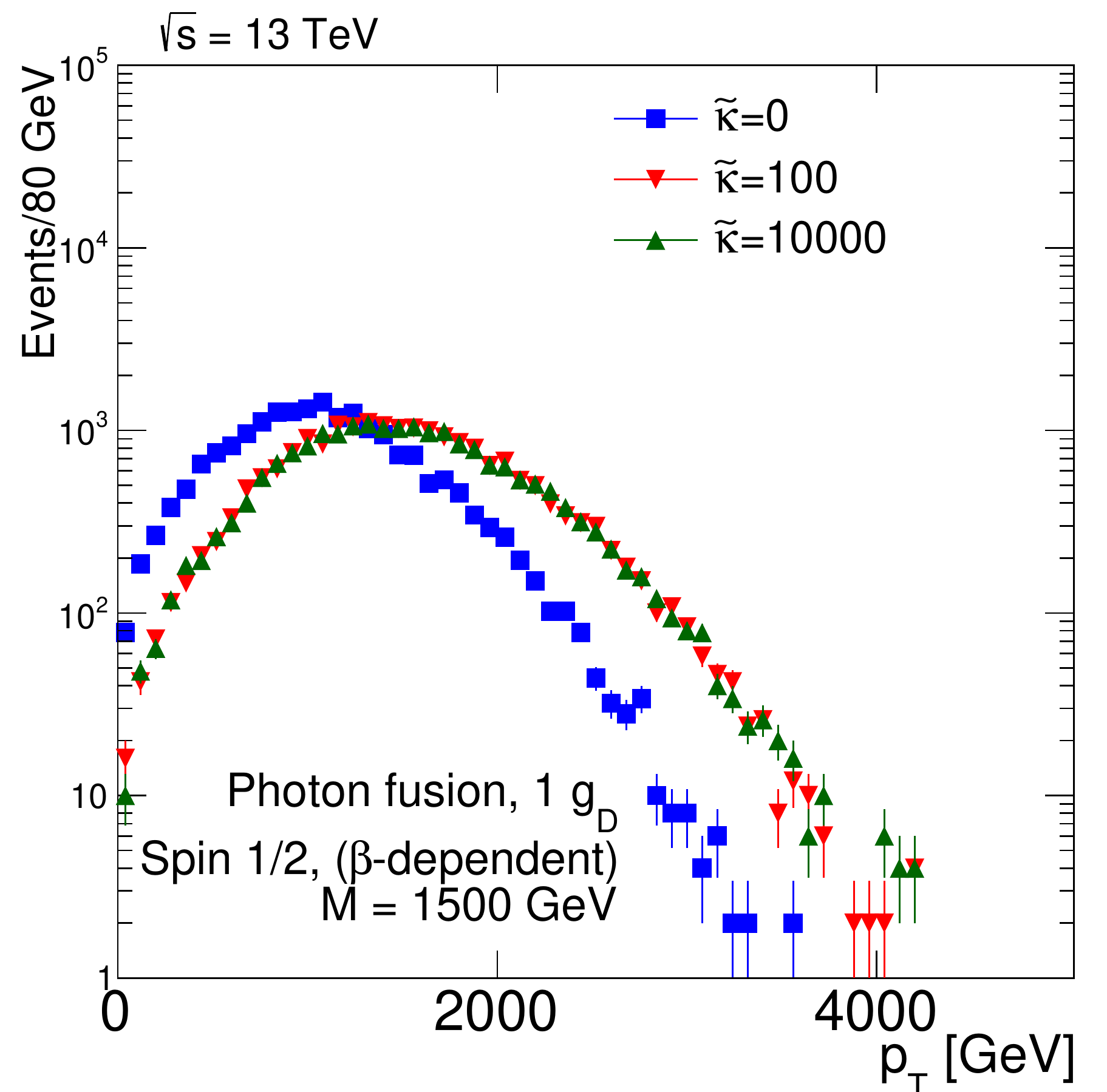}\hfill
\includegraphics[width=0.33\textwidth]{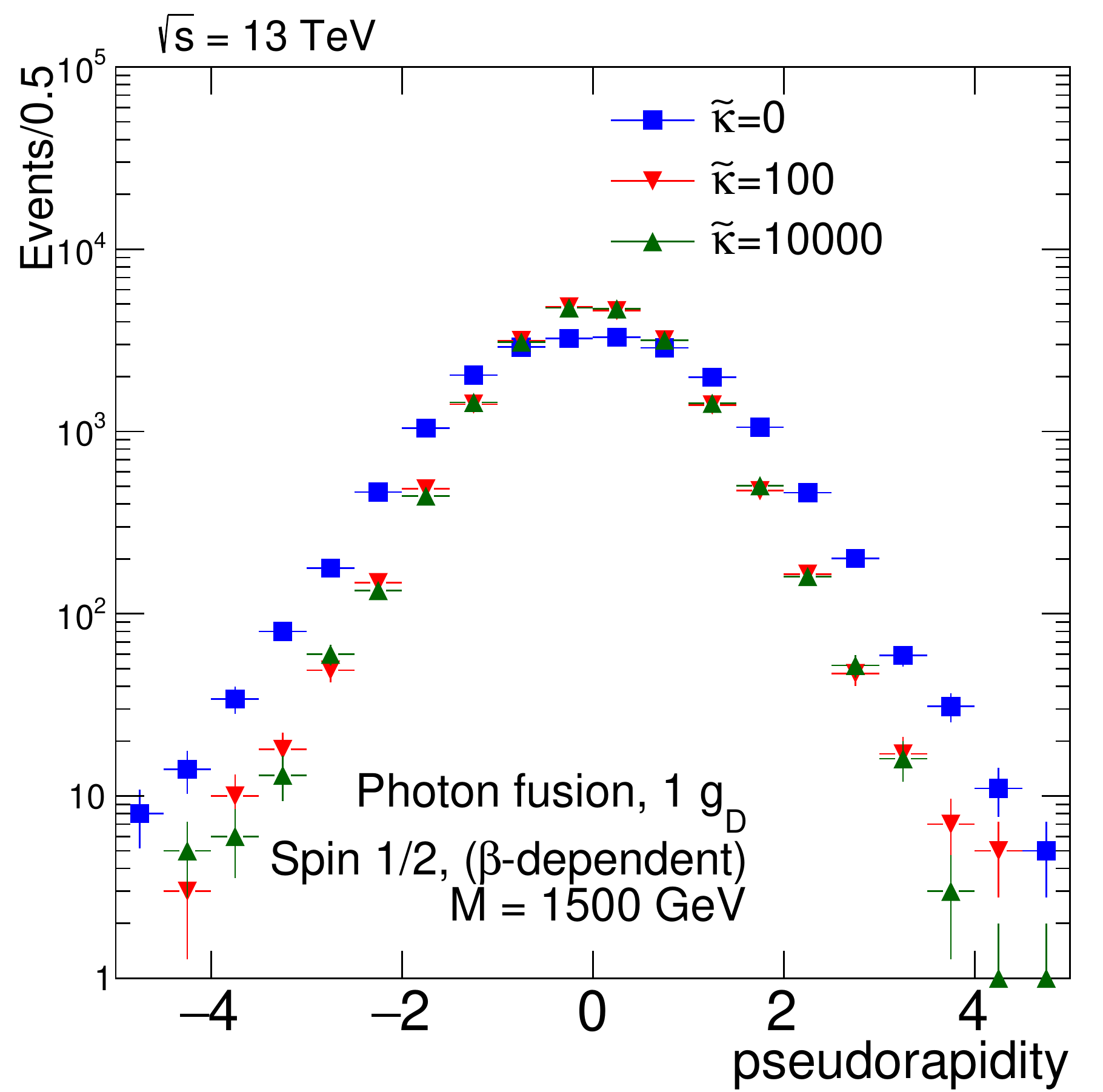}
\caption{Photon-fusion production at $\sqrt{s}=13$~TeV $pp$ collisions for spin-\half monopole, $\beta$-dependent coupling and various values of the $\tilde{\kappa}$ parameter: cross section versus monopole mass (left); \pt distribution for $M=1500~{\rm GeV}$ (centre); and $\eta$ distribution for $M=1500~{\rm GeV}$ (right)~\cite{Baines:2018ltl}. }\label{fig:spin12-kappa}
\end{figure}

For a spin-\half monopole, the dimensionless parameter $\tilde{\kappa}$, defined in~\eqref{ktilde}, is varied from zero (the SM scenario) to 10,000 for $\gamma\gamma$ collisions at $\sqrt{\sgg}=13~{\rm TeV}$. As evident from fig.~\ref{fig:spin12-kappa} (left), where the cross sections are plotted for $pp$ collisions at $\sqrt{s}=13$~TeV, the cross-section of the photon fusion process for $\tilde{\kappa}=0$ is going to zero very fast as $\beta\rightarrow 0$. However for non-zero $\tilde{\kappa}$, it remains finite even if $\beta$ goes to zero, as expected. 

The central and right-hand-side plots of fig.~\ref{fig:spin12-kappa} compare the \pt and $\eta$ distributions, respectively, between the SM value $\tilde{\kappa}=0$ and much higher values up to  $\tilde{\kappa}=10^4$. The SM-like case is characterised by a distinguishably ``softer'' \pt spectrum and a less central angular distribution than the large-$\tilde{\kappa}$ case. The latter case, on the other hand, seems to converge to a common shape for the kinematic variables as $\tilde{\kappa}$ increases to very large values. The common kinematics among large $\tilde{\kappa}$ values would greatly facilitate an experimental analysis targeting perturbatively reliable results. 

\begin{figure}[ht!]
\justify
\includegraphics[width=0.33\textwidth]{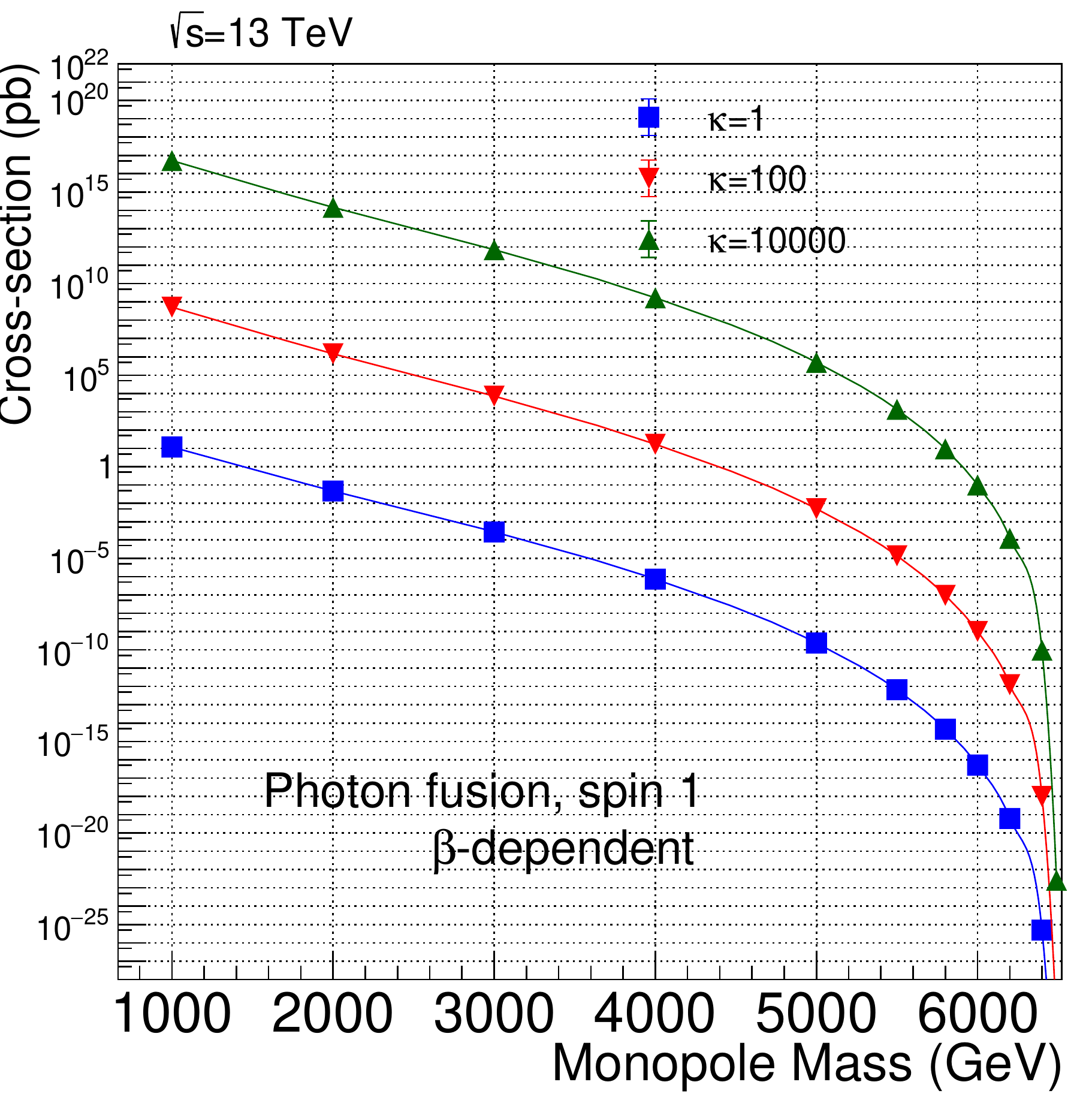}\hfill
\includegraphics[width=0.33\textwidth]{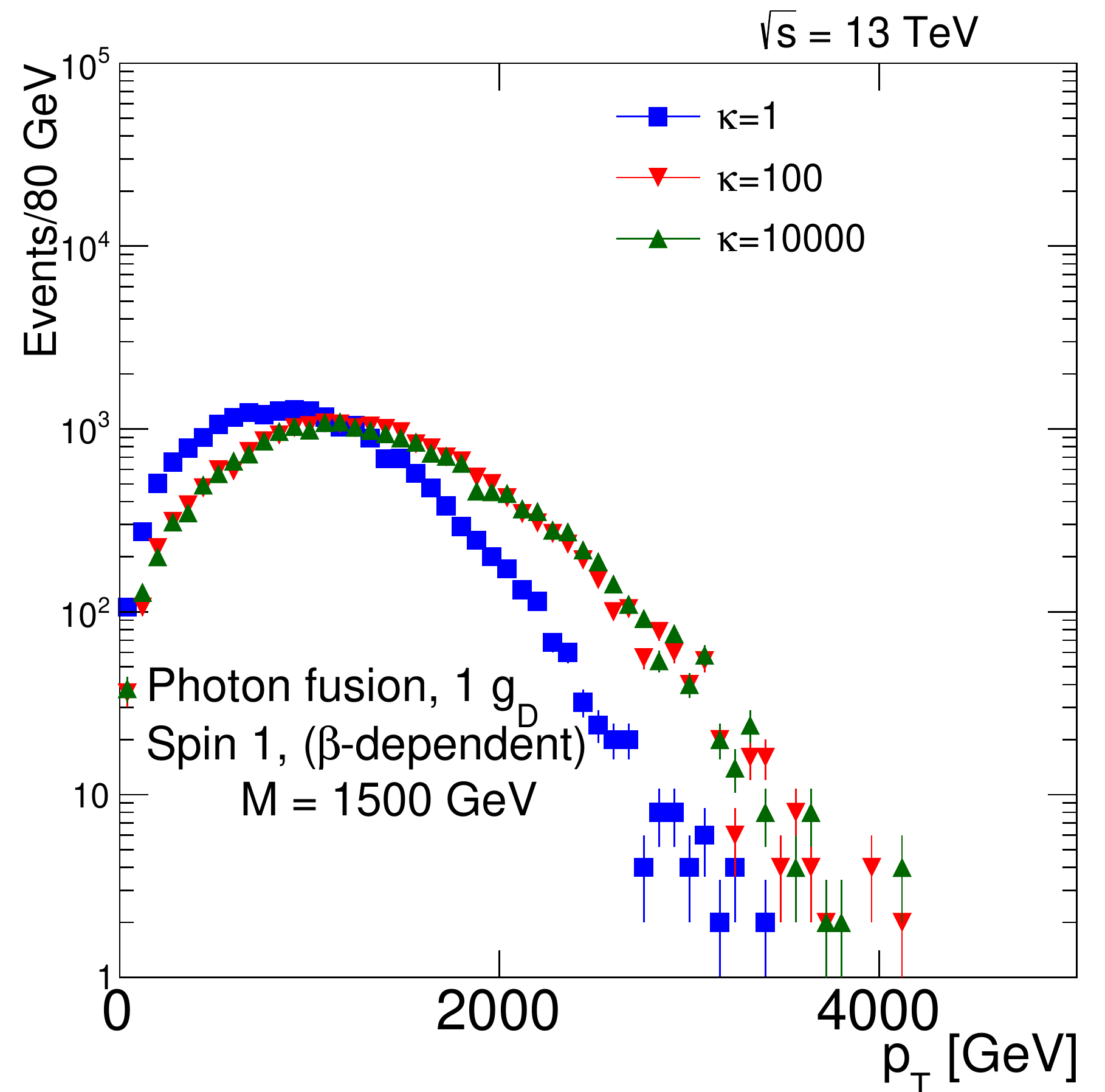}\hfill
\includegraphics[width=0.33\textwidth]{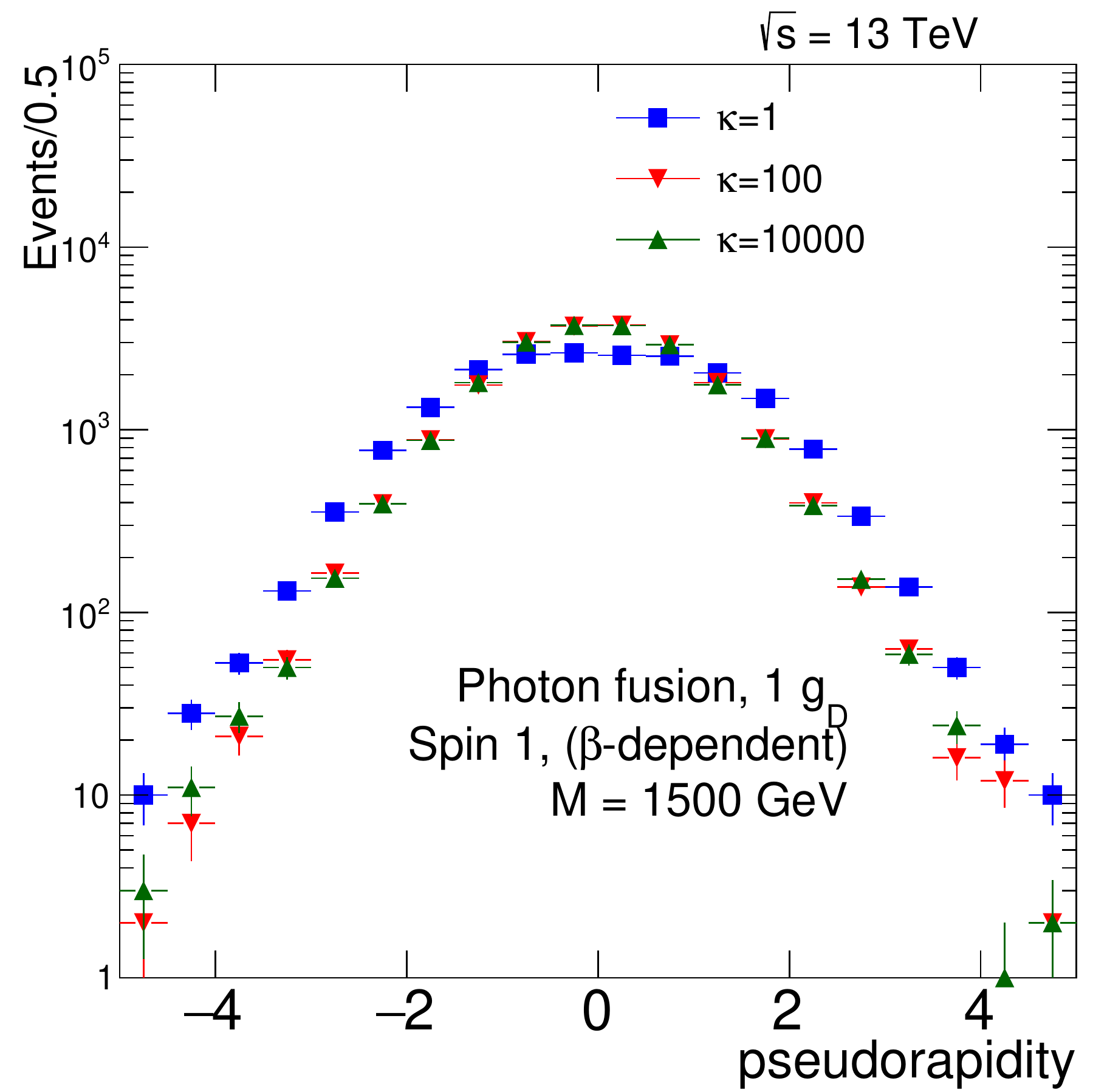}
\caption{Photon-fusion production at $\sqrt{s}=13$~TeV $pp$ collisions for spin-1 monopole, $\beta$-dependent coupling and various values of the $\kappa$ parameter: cross section versus monopole mass (left); \pt distribution for $M=1500~{\rm GeV}$ (centre); and $\eta$ distribution for $M=1500~{\rm GeV}$ (right)~\cite{Baines:2018ltl}.  }\label{fig:spin1-kappa}
\end{figure}

Repeating the same procedure this time for spin-1 monopoles, we vary the dimensionless parameter $\kappa$ from unity (the SM scenario) to 10,000 for the photon-fusion process. Similar to the spin-\half monopole case, the cross section for $\kappa=1$, i.e.\ the SM scenario, is going to zero very fast as $\beta\rightarrow 0$. However, for $\kappa>1$, the cross section becomes finite even if $\beta$ goes to 0. This observation remains valid when the cross section for $pp$ collisions, instead of $\gamma\gamma$ scattering, is considered. Indeed, fig.~\ref{fig:spin1-kappa} shows that for large values of $\kappa$ and $M \simeq \sqrt{s}/2$, which is equivalent to $\beta \ll 1$, the cross section, although very small, remains finite.

In fig.~\ref{fig:spin1-kappa}, a comparison of the \pt (centre) and $\eta$ (right) distributions, between the SM value $\kappa=1$ and much higher values up to  $\kappa=10^4$ is given for spin~1. As for spin~\half, the large-$\kappa$ curves converge to a single shape independent of the actual $\kappa$ value, with the SM-case yielding different distributions from the large-$\kappa$ case.  As discussed previously for fermionic monopoles, such an experimental analysis can be concentrated on the $\kappa$-dependence of the total cross section and the acceptance for very slow monopoles to provide perturbatively valid mass limits in case of non-observation of a monopole signal, since the kinematic distributions are almost $\kappa$-invariant in $pp$ collisions. The MoEDAL experiment~\cite{moedal-review}, in particular, being sensitive to slow monopoles can make the best out of this new approach in the interpretation of monopole-search results. 

\section{Conclusions \label{sec:concl}}

We dealt with the cross-section computation for pair production of monopoles of  spin $S=0,~\half,~1$,  through either photon-fusion or Drell--Yan processes. We have employed duality arguments to justify an effective monopole-velocity-dependent magnetic charge in monopole-matter scattering processes. Based on this, we conjecture that such $\beta$-dependent magnetic charges might also characterise monopole production. 

A magnetic-moment term proportional to a phenomenological parameter $\kappa$ is added to the effective Lagrangians for spins \half and~1. The lack of unitarity and renormalisability for an arbitrary  $\kappa$ value is not an issue, from an effective-field-theory point of view, given that the microscopic high-energy (ultraviolet) completion of the models considered above is unknown. In this sense, we consider the spin-1 monopole as a potentially viable phenomenological case worthy of further exploration, already applied in the recent monopole searches performed by MoEDAL~\cite{Acharya:2019vtb,moedalplb}. 

The possibility to use the parameter $\kappa$ in conjunction with the monopole velocity $\beta$ to achieve a perturbative treatment of the monopole-photon coupling is intriguing. Indeed, by limiting the discussion to very slow ($\beta \ll 1$) monopoles, the perturbativity is guaranteed, however, at the expense of a vanishing cross section. Nonetheless it turns out that the photon-fusion cross section remains finite \emph{and} the coupling is perturbative at the formal limits $\kappa\to\infty$ and $\beta\to 0$. This ascertainment opens up the possibility to interpret the cross-section bounds set in collider experiments, such as MoEDAL, in a proper way, thus yielding sensible monopole-mass limits. 

In addition, a complete implementation in \MAD of the monopole production is performed both for the photon-fusion and the Drell--Yan processes, also including the magnetic-moment terms. This tool allowed to probe for the first time at LHC the photon-fusion monopole production by the MoEDAL experiment~\cite{Acharya:2019vtb}. Furthermore, the experimental aspects of a perturbatively valid monopole search for large values of the magnetic-moment parameters and slow-moving monopoles have also been outlined, also in view of the corresponding kinematics.

\section*{Acknowledgements}
The author would like to thank the Corfu2018 organisers for the kind invitation to present this talk and her collaborators S.~Baines, N.~E.~Mavromatos, J.~L.~Pinfold and A.~Santra. This work was supported by the Generalitat Valenciana via a special grant for MoEDAL and via the Project PROMETEO-II/2017/033, by the Spanish MICIU via the grant FPA2015-65652-C4-1-R, by the Severo Ochoa Excellence Centre Project SEV-2014-0398, and by a 2017 Leonardo Grant for Researchers \& Cultural Creators, BBVA Foundation.


\end{document}